\documentclass[aps,prd,floatfix,noshowpacs,10pt,showkeys,tikz,nofootinbib,
nototlepage]{revtex4-2}

\usepackage[toc,page]{appendix}

\makeatletter
\newcommand*{\toccontents}{\@starttoc{toc}}
\makeatother

\usepackage[usenames,dvipsnames]{xcolor}

\definecolor{webgreen}{rgb}{0,0.75,0}
\definecolor{Mgreen}{rgb}{0,0.9,0.8}
\definecolor{webred}{rgb}{0.75,0,0}
\definecolor{webblue}{rgb}{0,0,0.75}
\definecolor{darkblue}{rgb}{0,0,0.7}
\definecolor{dunkelgrau}{rgb}{0.8,0.8,0.8}
\definecolor{lgray}{rgb}{0.95,0.95,0.95}
\definecolor{lgreen}{rgb}{0.95,1.00,0.90}
\definecolor{lblue}{rgb}{0.9,0.95,1.00}
\definecolor{lred}{rgb}{1.00,0.90,0.80}
\definecolor{shadecolor}{rgb}{1.00,0.92,0.82}

\DeclareMathAlphabet{\mathpzc}{OT1}{pzc}{m}{it}

\usepackage{graphicx}

\usepackage{natbib}

\usepackage{tikz-cd}
\usepackage{amssymb}
\usepackage{amsmath}

\tikzcdset{labels={font=\everymath\expandafter{\the\everymath\textstyle}}}

\usepackage{amsfonts}
\usepackage{mathrsfs}
\usepackage{bm}
\usepackage{color}

\usepackage[colorlinks=true,linkcolor=darkblue,citecolor=darkblue,urlcolor=darkblue,breaklinks=true]{hyperref}

\newcommand{\at}[2][]{#1|_{#2}}

\usepackage{orcidlink}

\makeatletter
\newcommand{\fmarki}{*}
\newcommand{\fmarkii}{\ensuremath{\dagger}}
\newcommand{\fmarkiii}{\ensuremath{\ddagger}}
\newcommand{\fmarkiv}{\ensuremath{\mathsection}}
\newcommand{\fmarkv}{\ensuremath{\mathparagraph}}
\newcommand{\fmarkvi}{\ensuremath{\|}}
\newcommand{\fmarkvii}{**}
\newcommand{\fmarkviii}{\ensuremath{\dagger\dagger}}
\newcommand{\fmarkix}{\ensuremath{\ddagger\ddagger}}

\def\@fnsymbol#1{{\ifcase#1\or \fmarki\or \fmarkii\or \fmarkiii\or \fmarkiv\or \fmarkv\or \fmarkvi\or \fmarkvii\or \fmarkviii\or \fmarkix \else\@ctrerr\fi}}
\makeatother

\renewcommand{\fmarki}{$\bigstar$}
\renewcommand{\fmarkii}{b$_2$}
\renewcommand{\fmarkiii}{c$_3$}
\renewcommand{\fmarkiv}{a$_4$}
\renewcommand{\fmarkv}{x$_5$}
\renewcommand{\fmarkix}{z$_9$}

\begin{document}



\vspace*{1cm}

\title{
Geometry of Classical Nambu-Goldstone Fields}

\author{Slobodan M. Rado\v sevi\' c \orcidlink{0000-0002-3211-1392}\;}
\email{slobodan@df.uns.ac.rs}


\affiliation{Department of Physics, Faculty of Sciences, University of Novi Sad, Trg Dositeja
 Obradovi\' ca 4, Novi Sad, Serbia\\
Department of Mathematics, University of Ja\' en, 
Ja\'en, Spain
}

\begin{abstract}

\noindent
A coordinate-free formulation of first order effective field theory, in which Nambu-Goldstone
fields are described as sections on associated bundle, is presented. This construction, which is based only on symmetry considerations, allows for a direct derivation of number and types of Nambu-Goldstone fields in a classical field theory without any reference to effective Lagrangian. A central role in classification is shown to be played by Lorentz-symmetry breaking order parameter which induces symplectic structure in the field space of the theory.

\end{abstract}






\maketitle  
\tableofcontents

\newpage
\section{Introduction}

\noindent Effective field theory (EFT) describes low energy sector
of physical systems which undergo spontaneous symmetry breakdown.
Being 
model independent, EFT
found applications across many areas of physics \cite{WeinbergQTF2,Burgess,BurgessBook,BraunerBook,Hofmann1,SciPostMagn} where
it not only provides a powerful tool for high-order perturbative calculations \cite{StrangeMass,Gasser,Gasser2,Hofmann3} but also
allows for a deeper insight into  physical problems \cite{WeinbergPast} or into the structure of other conventional methods used to deal with it   \cite{AnnPhys2015}.
The main dynamical degrees of freedom in EFT for internal symmetries are spin-zero Nambu-Goldstone (NG) bosons
and, in the case of Lorentz-invariant systems, they share  many common features -- they are  gapless excitations whose fields take values in the coset space $G/H$ and total number of which is
$\mathrm{dim}(G) - \mathrm{dim}(H)$, where $G$ represents the symmetry group of the action (or Hamiltonian) and $H$ is the symmetry group of the ground state. Also, their
dispersion relation is $\omega(\bm k ) \propto |\bm k|$.  On the other hand, situation becomes more involved
once the constraints  imposed  by the Lorentz symmetry are dropped -- 
neither the number of NG particles, nor their dispersion relation, are universal.
For example, $\mathrm{O}(3)$ ferromagnets and antiferromagnets share the same symmetry
breaking pattern, yet the spectrum of ferromagnets contains one NG boson with dispersion $\omega(\bm k) \propto \bm k^2$ while there are two NG bosons with $\omega(\bm k) \propto |\bm k|$ in the case of antiferromagnets. Also,  spinor Bose condensates support excitations with $\omega(\bm k) \propto \bm k^2$ and
$\omega(\bm k)  \propto |\bm k|$ at the same time, while explicit symmetry
breaking by finite charge density opens a gap for some of NG bosons.
Giving the obvious importance of these particles, efforts were made to understand the mechanism responsible for such a diversity of
NG bosons and their properties.

The original classification \cite{NielsChad} of bosonic excitations which accompany 
symmetry breakdown is based on the
form of their dispersion relations: excitations with
$\omega \propto |\bm k|^{2n+1}$ are labeled as type I, while those 
with $\omega \propto |\bm k|^{2n}$ as type II. If $N_{\rm{I}}$ and $N_{\rm{II}}$ denote the 
numbers of different flavors of 
type I and II bosons, respectively, the following inequality
holds: $N_{\rm{I}} + 2 N_{\rm{II}} \geq \mathrm{dim}(G/H)$.
This statement goes by the name of Nielsen-Chadha theorem (NCT).
Examples of type I NGB include pions in QCD, antiferromagnetic magnons as well as phonons in solids and superfluids,
while type II NGB appear
in ferromagnets and Bose-Einstein condensates \cite{Brauner,AnnRevCM}.
Even though NCT
leaves $n$ as an arbitrary integer, in the most cases of physical interest $n=0$ for type I and $n=1$ for type II NG bosons.

Subsequent development of nonrelativistic EFT came in the form of 
Lagrangian field theory  for NG fields
\cite{PRD}. 
By considering the symmetry conditions imposed on  the effective Lagrangian,
together with the commutation relations and
vacuum-expectation values of Noether charges  \cite{NambuJSP,Kaon}, Watanabe and Murayama
have traced the origin of two types of excitations to the structure of time
derivatives in the leading order term of the effective Lagrangian  \cite{JapanciPRL,PRX}.
The proof presented by Watanabe and Murayama is a perturbative one. That is,
they have obtained a general solutions of Leutwyler's equations in
local coordinate system near the origin of $G/H$. On the basis
of this solution,  the effective Lagrangian up to
quadratic order in NG fields is shown to, in general, contain coupled
canonically conjugate fields and their corresponding first time derivatives.
Accordingly, a new nomenclature for NG fields is proposed: NG fields which
are canonically paired in the effective Lagrangian are denoted
as type B, while remaining ones are denoted as type A.
Also, Watanabe, Murayama and Brauner \cite{JapanciPRL,PRX,WatanabeBraunerPRD}
showed that the numbers of type B  ($N_{\rm{B}}$) and type A ($N_{\rm{A}}$) NG boson flavors
are given by $N_{\rm{A}} = {\rm{dim}}(G/H) - \rm{rank} \rho$, 
$N_{\rm{B}} = \frac 12 \rm{rank} \rho$, where $\mathrm{i} V \rho_{ab} = \langle 0 | [Q_a, Q_b ] | 0 \rangle$, $V$ is the volume of system and
$\{ Q_a \}$ is the set of broken generators.
Since  the authors of \cite{PRX,WatanabeBraunerPRD} consider quantum theory
and discuss fluctuations and stability of the ground state,
their proof also implicitly makes use  of a special basis of $\mathfrak{g}$ chosen in such a way that only operators corresponding
to Cartan generators of $\mathfrak{g}$ have nonzero vacuum expectation values.
Another important aspect of the work by
Watanabe and Murayama is the geometric picture  underlining their
classification. As it is discussed in \cite{JapanciPRL,PRX}, 
the form of time derivatives in the leading order terms  essentially
determines  the presymplectic structure on $G/H$. Therefore, the 
coset space $G/H$ is   partially a phase space in which
canonically conugate NG fields (namely, type B bosons) live.
The identification of presymplectic structure on $G/H$ in these papers makes  use of the Borel's theorem which applies to
compact semisimple Lie groups and this puts some restrictions
on generality of their proof. 
However, the extension to noncompact group $G$
may be of interest when considering Higgs effective field theory,
spontaneous braking of Lorentz symmetry  or supergravity  \cite{ManoharPLB,ManoharJHEP,ProgressPNP,SpaceTimeSymBreak,SuperGravBook}. 
The classification was later refined to
  include  massive Nambu-Goldstone bosons
whose appearance accompanies explicit symmetry breakdown \cite{PRLMassive,PencoJHEP,HidakaPRD}.

Important constraints on the dynamics of NG fields, resulting from
the symmetry breakdown, can be studied within  classical field theory --
 a nonlinear $\sigma$-model with the target space
$G/H$. The dynamical degrees of freedom in classical theory
are sections on a properly chosen bundle and many of their features,
such as number and types of NG fields or their dispersion relation, are
shared between classical and lowest order quantum field theory. 
Also, classical field theory allows for a clear geometric formulation
which makes a sharp distinction between physical degrees of freedom (sections on a bundle) and local coordinates on $G/H$, which are devoid 
of any physical significance \cite{Bleecker,Fischer}. This is an important point since, giving the generality  EFT usually aims at \cite{WeinbergQTF2}, one should be in a position of deducing the properties of NG fields, such as their number and type, not from the
effective Lagrangian (which, by definition, is  a function of coordinates on the coset space), but from the properties of the system itself.
The main result of present
paper is that this can be done, at least on the level of classical description.
Also, the geometric arguments presented here allow for a unified 
description of massive and massless NG fields. 

The classical theory of NG fields   presented here is build around the
assumption of spontaneous  breakdown of  symmetry according to $G \to H$. We  assume that $G$ is connected, simply connected Lie group
and $H$, a symmetry
group of ground-state field configuration, a compact subgroup of $G$. To discuss two types of NG fields, we need to consider the existence of 
the Lorentz-symmetry breaking order parameter and its symmetry group to whom,
in general, $H$ is subgroup. As it turns out, Lorentz-symmetry breaking
order parameter is an element of $\mathfrak{g}^*$, a vector space
dual to the lie algebra of $G$, which carries the coadjoint representations
of Lie group $G$ and its Lie algebra, $\mathfrak{g}$. 
The role of $\mathfrak{g}^*$ turns out to be crucial in present construction,
as a distinguished element of $\mathfrak{g}^*$ (the order parameter) defines
the symplectic structure on space of its orbits under the  action of $G$ and
this  introduces two types of NG fields.
The three groups, which describe symmetries of the theory, are most conveniently taken into account 
starting from a principal $G$ bundle and  building the field space $\mathrm{E}$ as an associated bundle. Even though, in many applications, the field space
is a trivial bundle ($\mathrm{E} = (G/H) \times \mathrm{M}$, where $\mathrm{M}$ is spacetime),
the space $G/H$, as a bundle, is not, and the general theory of associated
bundles is well suited for this problem. Since the objective of  present paper concerns with types and number of NG fields, the field theory is presented using
1-jet formalism.
Finally, the associated bundle machinery    will allow us to formulate the theory of classical NG fields in a coordinate-free manner and to
derive classification of NG fields starting from the symmetry breaking 
pattern and properties of the order parameter, without any reference to
the effective Lagrangian or particular basis of $\mathfrak{g}$. 
We shall assume that are no redundancies
due to equivalent space-time symmetries \cite{PRLRedund, JHEPRedund}.

The Section \ref{NoetherJet} contains a brief review of classical field 
theory in the language of jets, together with the Noether theorem and momentum mapping, while the   symmetry breaking pattern, the Lorentz-symmetry breaking order parameter as an element of $\mathfrak{g}^*$ and fibration 
of $G/H$ are discussed in Section
\ref{SecSymBr}. A coordinate-free definition of type A and type B Nambu-Goldstone fields as sections on associated bundle build upon
three symmetry groups characterizing the system are presented in Section
\ref{NGFieldsAsocB}. Two types of coordinate systems on $G/H$, 
one of which allows for a decoupling of type A and type B NG fields
from each other  are described in Section \ref{CoordExpres}.
Finally, the case of explicit symmetry breaking, triggered by the chemical potential
coupled to the Noether charge in Hamiltonian, is discussed  Section \ref{ExplicitSSB},
while several examples of systems whose symmetry breaking patten induces several types of NG fields are given in Section 
\ref{ExamplesSect}. Some details on fiber bundles in general, and jet bundles in particular, as well as references
which deal rigorously with these objects, are given in Appendices \ref{AppA}and \ref{AppB}

\section{Noether symmetries and momentum mapping} \label{NoetherJet}

\noindent
Let $(\mathrm E, \pi_\mathrm{E}, \mathrm{M})$ represent a fiber bundle, where $\mathrm{M}$ denotes $m+1$
 dimensional spacetime,  $\mathrm F = \mathrm E_{\mathrm{p}}$, with $\mathrm{p} \in \mathrm{M}$,  
is $n$ dimensional typical fiber, $\mathrm E$ is $n+m+1$ dimensional total space
and $\pi_\mathrm{E}: \mathrm E \to \mathrm{M}$ is the projection. In this setting, 
we identify classical fields $s$ with
local sections of $\pi_\mathrm{E}$,   $s : \mathrm{M} \to \mathrm E$. Further,
let $(J^1 \pi_\mathrm{E}, \pi_1, \mathrm{M})$ denote the corresponding first jet bundle and $L \in C^\infty(J^1 \pi_\mathrm{E})$ be
the Lagrangian density. If $\mathrm{Vol} \mathrm{M}$ is a volume form on $ \mathrm{M}$,
the  action functional for  this classical field theory is 
\begin{equation}
S[s] = \int_{\mathrm U} (j^1 s)^* \mathcal{L},   \label{ActionJ1Def}
\end{equation}
where $\mathcal{L} = L \; \mathrm{Vol} \mathrm{M} \in \bigwedge_0^{m+1} \pi_1$ is the Lagrangian form, $\mathrm U \subset \mathrm{M}$ is 
a compact $m+1$ dimensional submanifold of $\mathrm{M}$ and $j^1 s$ is the first prolongation
of $s$. The classical field configurations are  extremals of the action (\ref{ActionJ1Def}),
i.e. those sections of $\pi_\mathrm{E}$ which satisfy corresponding Euler-Lagrange equations \cite{Saunders,AldayaGauge}.

Let now a Lie Group $G$ act on $\pi_\mathrm{E}$ via bundle automorphism
$(\eta, \bar{\eta})$  where $\eta: \mathrm E \to \mathrm E$ and $\bar{\eta}:\mathrm{M} \to \mathrm{M}$ is a diffeomorphism
such that $\pi_\mathrm{E} \circ \eta = \bar \eta \circ \pi_\mathrm{E}$ \cite{Momentum}. Under this action of $G$
the sections of $\pi_\mathrm{E}$ transform as
\begin{equation}
G:s \to s' \equiv \eta \cdot s := \eta \circ s \circ \bar{\eta}^{-1} . 
\end{equation}
Also, the first jet prolongation of $(\eta,\bar{\eta})$, to be denoted as $j^1(\eta,\bar{\eta}) \equiv j^1\eta : J^1\pi \to J^1 \pi$,
is defined to act on $j^1 s$ as
\begin{equation}
j^1s \to j^1 s' = j^1 \left( \eta \circ s \circ \bar{\eta}^{-1} \right):=
j^1 \eta \circ j^1 s \circ \bar{\eta}^{-1} .  \label{J1SectionDef}
\end{equation}
The actions of $(\eta, \bar \eta)$ and $j^1(\eta, \bar \eta)$ are conveniently represented in a commutative
diagram
\begin{equation}
 \begin{tikzcd}[row sep=2.0cm, column sep=2.5cm]
 &J^1\pi_\mathrm{E} \arrow[r, "j^1\eta"] \arrow[d, "\pi_{1,0}"]
 & J^1\pi_\mathrm{E}  \arrow[d, "\pi_{1,0}"']  \\
 &  \mathrm E \arrow[r, "\eta"]  \arrow[d, "\pi_\mathrm{E}"]  
 & \mathrm E \arrow[d, "\pi_\mathrm{E}"'] \\
 &  \mathrm{M} \arrow[r, "\bar{\eta}"] \arrow[u, bend left, "s" near end,start anchor={[xshift=-1.0ex]}]
   \arrow[uu, bend left, bend left=45, "j^1s",start anchor={[xshift=-1.2ex]}]
 & \mathrm{M} \arrow[u, bend right, "s'"' near end,start anchor={[xshift=+1.0ex]}]  
\arrow[uu, bend right, bend right=45, "j^1s'"',start anchor={[xshift=+1.2ex]}]
 \end{tikzcd} .  \label{j1fDef}
 \end{equation}
Here $(J^1\pi_\mathrm{E}, \pi_{1,0}, \mathrm E)$ is an affine bundle and $(j^1\eta, \eta)$ as well as $(j^1\eta, \bar \eta)$
are bundle automorphisms. 
If $X \in \chi (\mathrm E)$ is a projectable vector field with the flow $\eta_\epsilon$, 
whose compact support lies within $\pi_\mathrm{E}^{-1}(\mathrm U)$ 
\begin{equation}
X = \frac{\mathrm{d}}{\mathrm{d} \epsilon} \at[\bigg]{0} \eta_\epsilon, \hspace{1cm}
X\at[\bigg]{\pi_\mathrm{E}^{-1}(\partial \mathrm U)} = 0, \label{ProjXDef}
\end{equation}
then the pair $(\eta_\epsilon, \bar \eta_\epsilon)$ defines a bundle morphism for each 
$\epsilon$ and $\bar \eta_\epsilon$ denotes the flow of $\bar{X}$
defined by $\bar{X} \circ \pi_\mathrm{E} = \pi_{\mathrm{E} *}\circ X \in \chi (\mathrm{M})$ \cite{Momentum,SCFT}. We also note that the 
 Lie algebra of induced vector fields on $\mathrm E$ is homomorphic 
to $\mathfrak{g}$, the Lie algebra of group $G$. That is,  induced vector fields 
represent the Lie algebra $\mathfrak{g}$ and describe infinitesimal action of $G$ 
on various geometric objects on $\mathrm E$. When it becomes important, we shall 
denote by $X_{\bm A}$ the vector field on $\mathrm{E}$ induced by the action of
$\mathrm{exp} \bm A  \in G$, with $\bm A \in \mathfrak g$.

The variation of a 
section $s$, induced by the action of $G$, is a family of sections
$s_\epsilon := \eta_\epsilon \circ s \circ (\bar{\eta}_\epsilon)^{-1}$. Since the 
field $X$ is assumed to vanish at $\partial \mathrm U$, the variation $s_\epsilon$ 
coincides with $s$ at all points of $\partial \mathrm U$.
The invariance of action  (\ref{ActionJ1Def}) under the bundle automorphimsms 
may now be  expressed as \cite{Saunders}
\begin{equation}
0 = \frac{\mathrm{d}}{\mathrm{d} \epsilon} \at[\bigg]{0} \int_{\bar{\eta}_\epsilon
	(\mathrm U)} 
(j^1s_\epsilon)^* \mathcal{L}= \int_{\mathrm U} (j^1 s)^* \mathrm{d}_{X^1} \mathcal{L}, \label{ActJ1Inv}
\end{equation}
where we have used  (\ref{J1SectionDef}) and
properties of the pull-back operation. Here $\mathrm{d}_{X^1}$ denotes the 
Lie derivative with respect to the 1-jet prolongation of $X$,
\begin{equation}
X^1 = \frac{\mathrm{d}}{\mathrm{d} \epsilon} \at[\bigg]{0} j^1 \eta_\epsilon \label{X1Jet}
\end{equation}
and the condition $\mathrm{d}_{X^1} \mathcal{L}= 0$ simply states that  invariance of the Lagrangian form
under  transformations induced by the vector field $X$ always implies the 
invariance of action (\ref{ActionJ1Def}).
This constraint on $\mathcal{L}$, which
guarantees the invariance of classical equations of motion, 
may in fact be relaxed. If $x^\mu$ are (local) coordinates on $U \subset \mathrm{M}$ ($\mu = 0,1,2, \dots, m$) 
and $\mathrm{Vol_\mu M} = \mathrm{i}_{\partial_\mu} \mathrm{Vol} \mathrm{M}$,
where $\mathrm{i}$ denotes the interior product, the classical field configurations will  remain the same if
\begin{equation}
\mathrm{d}_{X^1} \mathcal{L}= \mathrm{d} \left[ \Delta^\mu \mathrm{Vol}_\mu \mathrm{M} \right]
\equiv \mathrm{d} \Delta, \label{ActionInvDiv}
\end{equation}
where $\Delta^\mu$ are functions
on $\mathrm{E}$ and $\mathrm{d} $ is the exterior derivative on $\mathrm{E}$. Indeed, a direct application of the Stokes theorem gives
\begin{equation}
\int_{\mathrm U} (j^1s)^* \mathrm{d} \left[ \Delta^\mu \mathrm{Vol}_\mu \mathrm{M} \right]
= \int_{\mathrm U}  \mathrm{d} \left[ (j^1s)^* \Delta^\mu \mathrm{Vol}_\mu \mathrm{M} \right]
= \int_{\partial \mathrm U} (j^1s)^* \Delta^\mu \mathrm{Vol}_\mu \mathrm{M}
\end{equation}
which obviously does not affect the field configurations in the interior of $\mathrm U$.

To discuss further implications of symmetry in a classical field theory, let us recall the definition of  Cartan form \cite{SCFT, Momentum, Saunders}.
In the case of first-order Lagrangian, 
the Cartan form is unique and is given by
\begin{equation}
\Theta_{\mathcal{L}} = \mathcal{L}+ \mathrm{d}_S(L) \label{CartanFormDef}
\end{equation}
where $S$ denotes the vertical endomorphism associated to $\rm{VolM}$, a 
vector-valued $m+1$ form on $J^1 \pi$
and $\mathrm{d}_S$ is the corresponding derivative operator \cite{Saunders}.
If $(x^\mu, u^a)$ are adapted coordinates on $\mathrm{E}$ ($a = 1,2 \dots, n$)
and $u^a_\mu$ are associated derivative coordinates,  then
\begin{equation}
\Theta_\mathcal{L}= \mathcal{L}+ \frac{\partial L}{\partial u^a_\mu} \theta^a \wedge \mathrm{Vol}_\mu \mathrm{M}  \label{CartanFormCoord}
\end{equation}
where $\theta^a$  are basis contact forms, defined by   $(j^1 s)^* \theta^a = 0$ and $s$ is an arbitrary section
on $\pi_\mathrm{E}$.
In local coordinates
\begin{equation}
\theta^a = \mathrm{d} u^a - u^a_\mu \mathrm{d} x^\mu.  \label{ContactFormCoord}
\end{equation}
Therefore, $(j^1 s)^* \mathcal{L}=  (j^1 s)^* \Theta_\mathcal{L}$. A section $s:\mathrm{M} \to \mathrm{E}$ is a solution of the
 Euler-Lagrange equations iff \cite{AldayaGauge}
\begin{equation}
(j^1 s)^*\left[ \mathrm{i}_Y \mathrm{d} \Theta_{\mathcal{L}}  \right] = 0  \label{ELECartan}
\end{equation}
for every vector field $Y$ on $J^1 \pi$. In particular, these equation remain 
the same if \cite{JMPCartan}
\begin{equation}
\mathrm{d}_{X^{1}} \Theta_\mathcal{L}=   \mathrm{d} \Delta, \label{CartInvDiv}
\end{equation}
with $\Delta$ as in (\ref{ActionInvDiv}). A vector field $X$, whose 1-jet extension $X^1$ generates the flow $j^1\eta_\epsilon$ which changes
the Cartan form  by a total divergence, 
is said to be a Noether symmetry \cite{SCFT}. Given the Noether symmetry condition on $\Theta_\mathcal{L}$,
we  introduce the covariant  momentum map in the Lagrangian representation 
\begin{equation}
\bm J^\mathcal{L}: J^1 \pi_\mathrm{E} \to \mathfrak{g}^* \otimes \wedge^n J^1 \pi_\mathrm{E},
\end{equation}
by defining
\begin{equation}
J^\mathcal{L}(\bm A) := \Delta_{\bm A} - \mathrm{i}_{X_{\bm{A}}^1} \Theta_\mathcal{L}, \hspace{1cm} \mathrm{with} \;\;\;\;\bm A \in \mathfrak{g},  \label{JADef}
\end{equation}
to be a $n$ form on $J^1\pi_\mathrm{E}$ given by $\langle  \bm J^\mathcal{L}(\mathrm{f}), \bm A   \rangle
:= J^\mathcal{L}(\bm A)|_{\mathrm{f}}$ for $\mathrm f \in J^1\pi_\mathrm{E}$. 
Above, $\mathfrak{g}^*$ denotes the dual of the Lie algebra $\mathfrak{g}$,
 $X_{\bm A}^1$ is the first jet prolongation of the vector field $X_{\bm A}$,
$\langle\;\;,\;\; \rangle$ stands for the natural pairing between the elements of $\mathfrak{g}$ and $\mathfrak{g}^*$
while $\Delta_{\bm A}$ is an $n$ form on $\mathrm{M}$ linear in $\bm A$.
If  $\{ \bm T_{i} \}$ with $i=1, 2 \dots, \mathrm{dim}(G)$, denotes a
basis of $\mathfrak{g}$ and $\{ \bm T^{*i} \}$ is the corresponding dual basis, we have
\begin{equation}
\bm J^\mathcal{L}= \bm T^{*i} \otimes J^\mathcal{L}_i  , \hspace{1cm} \mathrm{with} \;\;\;\; J^\mathcal{L}_i := J^\mathcal{L}(\bm T_{i}) = \Delta_i - \mathrm{i}_{X_{i}^1} \Theta_\mathcal{L}
\end{equation}
where
\begin{equation}
 X_i := X_{\bm{T}_i}
\end{equation}
are induced vector fields corresponding to the basis elements of $\mathfrak{g}$.
By using definition (\ref{JADef}) and the Noether condition (\ref{CartInvDiv}),
one easily shows that the quantity $\mathrm{d} J^\mathcal{L}(\bm A)$ is conserved
on configurations which satisfy EL equations
\begin{equation}
(j^1 s)^*\left[\mathrm{d} J^\mathcal{L}(\bm A)\right] = 0,  \label{NoetherMoment}
\end{equation}
which is just the Noether's first theorem. 
To summarize, if the action of the theory enjoys symmetry
described by the Lie Group $G$, which acts on fields by the
bundle automorphism, the corresponding conservation law may be
expressed in terms of  momentum map and  elements of $\mathfrak{g}$ (\ref{NoetherMoment}). The importance of this construction,
 at least for the purpose of the present paper, lies in the clear geometric
formulation which 
reveals the significance of $\mathfrak{g}^*$ for
underlying effective field theory.


\section{Symmetry breaking and coadjoint representation} \label{SecSymBr}

\noindent
Let us suppose now that the classical field theory we are interested in
describes a system which undergoes spontaneous symmetry breakdown according
to the scheme $G \to H$, where $G$ is connected, simply connected Lie group
and $H$ closed subgroup of $G$.
The low energy sector of such a theory is dominated by Nambu-Goldstone fields which take their values in the homogeneous space $\mathrm{E}_{\mathrm{p}} = G/H$. Each point of $\mathrm{a} \in G/H$ is
of the form  $gH = \{gh |h \in H  \}$ and corresponds to
a field configuration.
This homogeneous
space is equipped with a left transitive action $G \times G/H \to G/H$,
which we denote by $(g,\mathrm{a}) \to l(g,  \mathrm{a})$.
It is convenient to pick $eH \equiv O$, where $e$ denotes the unit
element in $g$, as the origin in $G/H$. The ground state field configuration
corresponds to $O$ and it is invariant under the action of $H$.
Then, standard arguments \cite{Grk} show that $T_O(G/H) \cong \mathfrak{g} /\mathfrak{h}$.
Further,  any homogeneous space admits a $G$ invariant metric and 
we may decompose $\mathfrak{g}$ into the direct sum \cite{Grk,Kowalski,KMS}
\begin{equation}
\mathfrak{g} = \mathfrak{h} \oplus \mathfrak{m}  \label{MDef}
\end{equation}
where $\mathfrak{m}$ is subspace of $\mathfrak{g}$, canonically isomorphic to
$T_O(G/H)$, invariant under adjoint action of $H$: 
$\mathrm{Ad}(H) \mathfrak{m} \subset \mathfrak{m}$.
Given a group $G$ and homogeneous space $G/H$, we may introduce
projection $\widetilde \pi_{\mathrm{P}}: G \to G/H, g \mapsto gH$. In this manner,
Lie group $G$ can be viewed as a principal  bundle with the base space $G/H$
and  typical fiber  $H$.

Of special importance is a situation in which the 
conserved quantity from the Noether theorem plays the role of the
order parameter.
Asides from breaking the Lorentz symmetry on a fundamental level,
such an occurrence puts further restrictions on the total space $\mathrm{E}$
and, consequently, classical field
configurations.
According to (\ref{NoetherMoment}), a Lorentz-symmetry breaking  order parameter  induces
a distinguished element of $\mathfrak{g}^*$, which we formally write as
\begin{equation}
\bm \Sigma :=  \bm{T}^{*i}  \left \langle  \frac{1}{V} \;Q_i \right\rangle_{\rm{g.s.}}   =\; \bm{T}^{*i} \left\langle  \frac{1}{V}\int_{\bar{\mathrm{U}}}(j^1s)^*\left[ \Delta_i^0 - X^0_i L- X^a_i \frac{\partial L}{\partial u^a_0}
+ X^0_i \frac{\partial L}{\partial u^a_\nu} u^a_\nu   \right] \mathrm{Vol_0 M} \right\rangle_{\rm{g.s.}} \equiv \Sigma_i \bm{T}^{*i} \label{SigmaAverage}
\end{equation}
where  $Q_i$ is a Noether charge,   $X^\mu_i, X^a_i$ are the components of $X_i$ in a local chart on $\pi_\mathrm{E}$, $\langle \dots \rangle_{\rm{g.s.}}$ denotes
the ground state average value of a corresponding quantum (statistical) theory, $\bar{\mathrm{U}}$ is a spacelike region of $\mathrm{M}$ and $V$ is the volume of $\bar{\mathrm{U}}$. 
While it is obvious that this average cannot be calculated within classical theory, this is of no concern to us and we shall simply
assume that $\Sigma_i \neq 0$ break the Lorentz invariance.
The order parameter $\bm \Sigma$ transforms with the coadjoint representation $\mathrm{Ad}^*$, defined by
$\langle \mathrm{Ad}^*(g)\bm \Sigma, \bm A \rangle = \langle \bm \Sigma,  \mathrm{Ad}(g^{-1}) \bm A  \rangle$, for $\bm A \in \mathfrak{g}$ \cite{Kirillov}. Further, let us introduce the coadjoint orbit as the
set ${\mathcal{O}} = \{ \mathrm{Ad}^*(g) \bm \Sigma | g \in G$ \}. 
Since all the points in ${\mathcal{O}}$
are generated by the action of $G$, this set  may be identified with the
homogeneous space $G/\mathrm{Stab}(\bm \Sigma)$, where  $\mathrm{Stab}(\bm \Sigma)$ denotes the
isotropy group of $\bm \Sigma$
\begin{equation}
\mathrm{Stab}(\bm \Sigma) = \{ g \in G | \mathrm{Ad}^*(g) \bm \Sigma = \bm \Sigma \}.
\end{equation}
Therefore, the group $G$ may also be viewed as the principal fiber bundle with projection $\pi_{\tilde{\mathrm{P}}}$,
base space $G/\mathrm{Stab}(\bm \Sigma)$ and typical fiber $\mathrm{Stab}(\bm \Sigma)$. 
In complete analogy with (\ref{MDef}), we may decompose $\mathfrak{g}$ also as
\begin{equation}
\mathfrak{g} = \mathfrak{stab}(\bm \Sigma) \oplus \mathfrak{n}   \label{Ndef}
\end{equation}
where 
$\mathfrak{n}  \cong \mathfrak{g} /\mathfrak{stab}(\bm \Sigma) \cong T_{\tilde O}(G/\mathrm{Stab}(\bm \Sigma))$,  $\tilde{O}$ is the origin of $\mathcal{O}$ and 
\begin{equation}
\mathfrak{stab}(\bm \Sigma)  = \{ \bm B \in \mathfrak{g} | \mathrm{ad}^*(\bm B) \bm \Sigma = 0   \}, \hspace{1cm} 
\langle \mathrm{ad}^*(\bm B) \bm \Sigma, \bm A \rangle = \langle \bm \Sigma, - \mathrm{ad}(\bm B) \bm A \rangle  \label{StabSigmaAlg}
\end{equation}
is the Lie algebra of $\mathrm{Stab}(\bm \Sigma)$ which acts on the order parameter by corresponding coadjoint representation $\mathrm{ad}^*$.
The order parameter $\bm \Sigma$ must also be $H$ invariant and, therefore, in general
$H \subseteq \mathrm{Stab}(\bm \Sigma)$. We shall assume that  $H$ as well as $\mathrm{Stab}(\bm \Sigma)$ are compact subgroups of $G$.

When $H \subset \mathrm{Stab}(\bm \Sigma)$, and both $H$ and $\mathrm{Stab}(\bm \Sigma)$ are compact subgroups of $G$, previously
introduced fibrations induce yet another one. With the natural projection
$f: G/H \to G/\mathrm{Stab}(\bm \Sigma)$, given by $gH \mapsto g \mathrm{Stab}(\bm \Sigma)$,
homogeneous space $G/H$ becomes a  bundle itself \cite{Besse}: the base space is
$G/\mathrm{Stab}(\bm \Sigma)$ and the typical fiber is  $\mathrm{Stab}(\bm \Sigma)/H$. 
The pattern of spontaneous symmetry breaking (SSB), together with the structure of order parameter configuration,
completely determines three fibrations summarized in the diagram:
\begin{equation*}
	\begin{tikzcd}[row sep=huge]
		&
		G  \arrow[dl,swap,"\widetilde{\pi}_\mathrm{P}"] \arrow[dr,"\pi_{\tilde{\mathrm{P}}}"] &
		\\
		G/H \arrow[rr,"{f}"] & & G/\mathrm{Stab}(\bm \Sigma) .
	\end{tikzcd}
\end{equation*}
Of particular interest in the following  is the fact that ${\mathcal{O}}= G/\mathrm{Stab}(\bm \Sigma) $
is a symplectitc manifold equipped with the Kirillov-Kostant symplectic form \cite{Kirillov,GuerreroJMP,PhysRevResearch}. Thus, a $G$ invariant Lagrangian, together with
$\mathrm{Stab}(\bm \Sigma) \subset G$ invariant Lorentz-symmetry breaking order parameter and $H \subset \mathrm{Stab}(\bm \Sigma) \subset G$
invariant ground state field configuration, naturally lead to 
two types of NG fields: the ones who take values in a symplectic manifold 
$G/\mathrm{Stab}(\bm \Sigma)$ and the rest whose values belong to
the space $\mathrm{Stab}(\bm \Sigma)/H$.
Indeed, these are type B and type A Nambu-Goldstone fields introduced in \cite{JapanciPRL}. If $H=\mathrm{Stab}(\bm \Sigma)$, one takes
symmetry breaking pattern to be $G \to \mathrm{Stab}(\bm \Sigma)$ and constructs a theory of type B NG fields [See \ref{HeisExample} for an example] whenever 
$G$ is non-Abelian \cite{PRD}.

\section{Nambu-Goldstone fields on associated bundle}  \label{NGFieldsAsocB}

\noindent As we saw in previous section, a Lorentz-symmetry 
breaking order parameter naturally induces the fibration of $G/H$
so that Nambu-Goldstone fields  come into two types. In
this section we shall explicitly construct the bundle
$(\mathrm{E}, \pi_\mathrm{E}, \mathrm{M})$, with $\mathrm{E}_{\mathrm p} = G/H$
and show that both types of NG fields  may be defined
in a coordinate-free manner, in a relatively general case. To do so, we first recall some
basic facts about associated bundles \cite{KMS} and then proceed with the
exposition.


\subsection{Associated bundles and maps}

\noindent To construct the fiber bundle $(\mathrm{E}, \pi_\mathrm{E}, \mathrm{M})$,
where $\mathrm{M}$ is spacetime and typical fiber is $G/H$, we 
start from the principal $G$ bundle $(\mathrm{P}, \pi_{\mathrm{P}}, M)$ equipped with right principal action $\mathrm{P} \ni \mathrm{u} \mapsto \mathrm{u} \cdot g$  and pick
a local section $\sigma: \mathrm{M} \to \mathrm{P}$ to provide a local trivialization.
If we consider a reduction of symmetry group $G \to H$, then there
is a well defined associated bundle $\mathrm{P}_G[G/H;l]$ with the fiber
$G/H$  and the left action $l: G \times (G/H) \to G/H$.
The total space of associated bundle $\pi_\mathrm{E}$ is build  using projection
$q: \mathrm{P} \times (G/H) \to \mathrm{E}$, which is defined as an equivalence class
with respect to orbits of the right action
$R:(\mathrm{P} \times (G/H)) \times G \to \mathrm{P} \times (G/H)$, defined by
$R\left( (\mathrm{u},\mathrm{a}), g  \right) = \bigl(\mathrm{u} \cdot g, l(g^{-1}, \mathrm{a})\bigr)$ where dot denotes
the right principal action. If we fix $\mathrm{u} \in \mathrm{P}$, then 
$q_\mathrm{u}: \{\mathrm{u}\} \times (G/H) \to \mathrm{E}_{\pi_{\mathrm{P}}(\mathrm{u})}$ is a diffeomorphism
for each $\mathrm{u} \in \mathrm{P}$. The second important map defined for an associated bundle $\pi_\mathrm{E}$ is $\tau: \mathrm{P} \times_\mathrm{M} \mathrm{E} \to G/H$, where
$\tau (\mathrm{u}_\mathrm{p}, \mathrm{a}_\mathrm{p}): = q^{-1}_{\mathrm{u}_\mathrm{p}}(\mathrm{a}_\mathrm{p})$ with $\pi(\mathrm{a}_\mathrm{p}) = \pi(\mathrm{u}_\mathrm{p}) = \mathrm{p} \in \mathrm{M}$. The maps $q$ and
$\tau$ can be used to express a local section   $s$ of $\pi_\mathrm{E}$ in terms of 
the frame form $\bar{s}: \mathrm{P} \to G/H$ and a local section of $\pi_\mathrm{P}$ as
\begin{equation}
	s = q \circ(\mathrm{i} \mathrm{d}_\mathrm{P} \times \bar s\bigr) \circ \sigma = 
	q\bigl(\sigma, \bar s \circ \sigma\bigr) 
\end{equation}
where
\begin{equation}
	\bar s = \tau \bigl(\mathrm{i} \mathrm{d}_\mathrm{P} \times s \circ \pi_\mathrm{P}\bigr).
\end{equation}
Various maps introduced here are conveniently represented in a commutative diagram 
\begin{equation} \hspace{-1.0cm}
	\begin{tikzcd}[row sep=1.8cm, column sep=2.0cm]
& G/H \arrow[r, leftarrow, "\tau"] 	& \mathrm{P}\times_{\mathrm{M}} \mathrm{E} \arrow[r, leftarrow, "\mathrm{i} \mathrm{d}_{\mathrm{P}} \times s \circ \pi_{\mathrm{P}}"]	& 
\arrow[ll,  bend left = 15, "\bar{s}"]  \mathrm{P} \arrow[r, "\mathrm{i} \mathrm{d}_{\mathrm{P}} \times \bar{s}"] \arrow[d, "\pi_{\mathrm{P}}"'] & 
\mathrm{P} \times (G/H) 
		\arrow[r,"q"] & 
		\mathrm{E} \arrow[d, "\pi_{\mathrm{E}}"']
		\\
& \mathrm{M} \arrow[u, "\phi"'] \arrow[rr, leftarrow, "\mathrm{i} \mathrm{d}_\mathrm{M}"] && \mathrm{M} \arrow[rr, "\mathrm{i} \mathrm{d}_\mathrm{M}"] \arrow[u, bend right, "\sigma"'] &&  \mathrm{M} \arrow[u, bend right,"s"'] &
	\end{tikzcd} \label{PrincEbundle}
\end{equation}
where $\mathrm{i} \mathrm{d}_\mathrm{M}$ denote the identity map on $\mathrm{M}$. 
The formalism of first-order jets outlined in Section \ref{NoetherJet} can now
be used to construct the classical NG fields (see recent papers  \cite{PRLJB,Arxiv2023}).
The map $\phi: \mathrm{M} \to G/H$ defined as
\begin{equation}
\phi = \bar s \circ \sigma = \tau \bigl(\mathrm{i} \mathrm{d}_\mathrm{P} \times s \circ \pi_\mathrm{P}\bigr) \circ \sigma = \tau \bigl( \sigma, s  \bigr)
\end{equation}
can be understood as a local representative of a section $s$ and  
$\phi$ is  usually the object which is considered as a classical Nambu-Goldstone field \cite{JapanciPRL,PRX,AnnPhys}. Given $\phi$,
we can reconstruct the section $s$ using  $q$ projection
\begin{equation}
s = q \bigl( \sigma, \phi   \bigr).   \label{SSectDef}
\end{equation}
The local representative $\phi$ has ${\rm{dim}}(G) - {\rm{dim}}(H)$ components,
and they are  split  into physical degrees of freedom
carried by NG fields of  type A and type B.

\subsection{Type B Nambu-Goldstone fields} \label{TypeBSubsect}

\noindent Suppose now that the ground state of system is such
that the Noether charge develops vacuum
expectation value which we identify with the order parameter $\bm \Sigma \in \mathfrak{g}^*$. The coset space $G/H$ may now be viewed as a bundle
$\bigl(G/H, f, G/\mathrm{Stab}(\bm \Sigma)\bigr)$, where $f$ denotes the
canonical projection $f: G/H \to G/\mathrm{Stab}(\bm \Sigma)$,
$gH \mapsto g\mathrm{Stab}(\bm \Sigma)$. The 
typical fiber of $f$ is $\mathrm{Stab}(\bm \Sigma)/H$ and we shall focus on this space in next subsection. 

The canonical projection $f$ is 
well defined and we may simply introduce type B Nambu-Goldstone fields
as local representatives of section $s:\mathrm{M} \to \mathrm{E}$ which take values in
$G/\mathrm{Stab}(\bm \Sigma)$. That is, for each ${\rm{a}} = \phi(\mathrm p) \in G/H$,
there is a map ${\color{red}\phi_{\rm B}}: \mathrm{M} \to G/\mathrm{Stab}(\bm \Sigma)$
\begin{eqnarray}
{\color{red} \phi_{\mathrm{B}}} & = & f \circ \phi = f \circ \bar s \circ \sigma \nonumber \\
& = & f \circ \tau \bigl(\mathrm{i} \mathrm{d}_\mathrm{P} \times  s \circ \pi_{\mathrm{P}}\bigr) \circ \sigma  \label{PhiBDef}
\end{eqnarray}
such that ${\color{red}\phi_{\rm B}}(\mathrm p) = f\bigl(  \phi(\mathrm p) \bigr)$.
This definition is given in the following diagram
\begin{equation} 
	\begin{tikzcd}[row sep=1.8cm, column sep=2.3cm]
		& \mathrm{P} \arrow[rr,  bend right = 15, "\bar{s}"']
		\arrow[d, "\pi_{\mathrm{P}}"]
		\arrow[r, "\mathrm{i} \mathrm{d}_{\mathrm{P}} \times s \circ \pi_{\mathrm{P}}"] 
		& \mathrm{P} \times_{\mathrm{M}} \mathrm{E}
		 \arrow[r, "\tau"] & G/H \arrow[r, "f"] & G/\mathrm{Stab}(\bm \Sigma) \\
		 & \mathrm{M} \arrow[rr, "\mathrm{i} \mathrm{d}_{\mathrm{M}}"] 
		 \arrow[u, bend left, "\sigma"]
		 && \mathrm{M} 
		 \arrow[u, "\phi"]
		 \arrow[r, "\mathrm{i} \mathrm{d}_{\mathrm{M}}"] 
		 & \mathrm{M} 
		 \arrow[u, red, "{\color{red}\phi_{\mathrm{B}}}"]
		 &
	\end{tikzcd}.
\end{equation}
As we remarked earlier, $G/\mathrm{Stab}(\bm \Sigma)$ is a symplectic
manifold. Thus, type B NG fields come in pairs and the number of physical
degrees of freedom which they carry is
\begin{equation}
N_{\mathrm B} = \frac 12 \mathrm{dim} \biggl( G/\mathrm{Stab}(\bm \Sigma)  \biggr)  \label{NPhiBDef}
\end{equation}
Of course, this is exactly the result obtained in 
\cite{JapanciPRL,PRX,HidakaPRL,Kapustin}.

\subsection{Type A Nambu-Goldstone fields}  \label{TypeASubsect}

\noindent To give a proper definition of type A Nambu-Goldstone fields,
recall that the bundle $\bigl(G/H, f, G/\mathrm{Stab}(\bm \Sigma)    \bigr)$
may be viewed as an associated bundle to the principal $\mathrm{Stab}(\bm \Sigma)$ bundle $\bigl(G \equiv \tilde{\mathrm{P}}, \pi_{\tilde{\mathrm{P}}}, G/\mathrm{Stab}(\bm \Sigma) \bigr)$ \cite{Besse}. Thus
\begin{equation}
\Bigl(G/H, f, G/\mathrm{Stab}(\bm \Sigma)    \Bigr) = \tilde{\mathrm{P}}_{\mathrm{Stab}(\bm \Sigma)} \Bigl[\mathrm{Stab}(\bm \Sigma)/H; \tilde l   \Bigr]
\end{equation}
where $\tilde l$ denotes the left action of $\mathrm{Stab}(\bm \Sigma)$
on ${\mathrm{Stab}(\bm \Sigma)}/H$. The associated bundle 
$\tilde{\mathrm{P}}_{\mathrm{Stab}(\bm \Sigma)} \bigl[\mathrm{Stab}(\bm \Sigma)/H; \tilde l   \bigr]$ comes equipped with mappings $\tilde q$ and $\tilde \tau$ which are defined analogously to  $q$ and $\tau$  of 
$\mathrm{P}_G[G/H;l]$.  We can use
$\tilde q$ to connect the total space of 
$\tilde{\mathrm{P}}_{\mathrm{Stab}(\bm \Sigma)} \bigl[\mathrm{Stab}(\bm \Sigma)/H; \tilde l   \bigr]$ with the fibers of $(\mathrm{E}, \pi_\mathrm{M}, \mathrm{M})$ and
 $\tilde \tau$ to define
NG fields which take values in $\mathrm{Stab}(\bm \Sigma)/H$.

Let $\tilde \sigma$ denote a local section on $\pi_{\tilde{\mathrm{P}}}$, and
$c$ is be local section on $\bigl( G/H, f, \tilde{\mathrm{M}} \equiv G/\mathrm{Stab}(\bm \Sigma)    \bigr)$ determined by 
$\phi = c \circ {\color{red}\phi_{\mathrm B}}$.
Since $f \circ \phi = {\color{red}\phi_{\mathrm B}}$, we have $f \circ c = \rm{id}_{\tilde{M}}$, so that $c$ is indeed a section on $\bigl( G/H, f, \tilde{\mathrm{M}} \equiv G/\mathrm{Stab}(\bm \Sigma)    \bigr)$.
If $\bar{c}$ denotes the 
frame form of $c$,
we have the following diagram
\begin{equation} \hspace{-1.5cm}
	\begin{tikzcd}[row sep=1.8cm, column sep=0.9cm]
		& \mathrm{Stab}(\bm \Sigma)/H \arrow[r, leftarrow, "\tilde \tau"] 	& 
		\tilde{\mathrm{P}}\times_{\tilde{\mathrm{M}}} \tilde{\mathrm{E}} \arrow[r, leftarrow, "\mathrm{i} \mathrm{d}_{\tilde{\mathrm{P}}} \times c \circ \pi_{\tilde{\mathrm{P}}}"]	& 
		\arrow[ll,  bend left = 15, "\bar{c}"]  \tilde{\mathrm{P}} \equiv G \arrow[r, "\mathrm{i} \mathrm{d}_{\mathrm{P}} \times \bar{c}"] \arrow[d, 
		"\pi_{\tilde{\mathrm{P}}}"'] & 
		\tilde{\mathrm{P}} \times (\mathrm{Stab}(\bm \Sigma)/H) 
		\arrow[r,"\tilde q"] & \tilde{\mathrm{E}} \equiv G/H 
		\arrow[d, "f"]
		\\
		& \tilde{\mathrm{M}}  \arrow[u, "\tilde \phi"'] \arrow[rr, leftarrow, "\mathrm{i} \mathrm{d}_{\tilde{\mathrm{M}}}"] && \tilde{\mathrm{M}} \equiv G/\mathrm{Stab}(\bm \Sigma) \arrow[rr, "\mathrm{i} \mathrm{d}_{\tilde{\mathrm{M}}}"] \arrow[u, bend right = 30, "\tilde \sigma"'] && \tilde{\mathrm{M}} \arrow[u, bend left, "c"] &\\
		&\mathrm{M} \arrow[rr, leftarrow, "\mathrm{i} \mathrm{d}_\mathrm{M}"] \arrow[u, red, "{\color{red}{\phi_{\mathrm{B}}}}"']
		\arrow[uu, bend left = 30, blue,"{\color{blue}\phi_{\mathrm A}}"] && \mathrm{M} \arrow[rr, "\mathrm{i} \mathrm{d}_\mathrm{M}"] 
		\arrow[u, red, "{\color{red}{\phi_{\mathrm{B}}}}"] && \mathrm{M}
		\arrow[u, red, "{\color{red}{\phi_{\mathrm{B}}}}"]
		\arrow[uu, bend right = 35, "\phi"']
	\end{tikzcd} \label{PrincEbundle2}
\end{equation}
and type A Nambu-Goldstone fields are defined as
\begin{eqnarray}
{\color{blue}\phi_{\mathrm A}} & = & \tilde{\phi}\circ  {\color{red}{\phi_{\mathrm{B}}}}
= \bar{c} \circ \tilde \sigma \circ 
{\color{red}{\phi_{\mathrm{B}}}} \nonumber \\
& = & \tilde \tau \bigl( \mathrm{i} \mathrm{d}_{\tilde{\mathrm{P}}} \times c \circ \pi_{\tilde{\mathrm{P}}} \bigr) \circ \tilde \sigma \circ 
f \circ \phi   \nonumber \\
& = & \tilde \tau \Bigl(  \tilde \sigma \circ {\color{red}{\phi_{\mathrm{B}}}}, \phi  \Bigr). \label{PhiADef}
\end{eqnarray}
In other words, given $\phi(\mathrm p) \in G/H$ and ${\color{red}\phi_{\rm B}}(\mathrm p) \in G/\mathrm{Stab}(\bm \Sigma)$, there is a corresponding point in $\mathrm{Stab}(\bm \Sigma)/H$ given by ${\color{blue}\phi_{\rm A}}(\mathrm p)$.
Since there is no natural symplectic
structure on $\mathrm{Stab}(\bm \Sigma)/H$, the number of physical degrees
freedom described by type A Nambu-Goldstone fields is
\begin{equation}
N_{\mathrm{A}} = \mathrm{dim} \Bigl(  \mathrm{Stab}(\bm \Sigma)/H  \Bigr).
\label{NPhiADef}
\end{equation}
Note tat this definition of type A NG fields assumes the existence
of Lorentz-symmetry breaking order parameter as, given $\phi(\mathrm p)$, ${\color{blue}\phi_{\mathrm A}}(\mathrm p)$ depends on ${\color{red}{\phi_{\mathrm{B}}}}(\mathrm p)$. In the case when 
$\bm \Sigma = \bm 0$, one proceeds in a usual way and introduces local
representatives of
type A NG fields as maps ${\color{blue}\phi_{\mathrm A}}: \mathrm{M} \to G/H$
\cite{AnnPhys} and
\begin{equation}
N = \mathrm{dim} \Bigl( G/H  \Bigr).
\end{equation}
is the total number of physical degrees of freedom in this case.

\subsection{Dynamical type A and B NG fields}

\noindent Results of \ref{TypeBSubsect} and \ref{TypeASubsect} show that given section $s: \mathrm{M} \to \mathrm{E}$, which is a solution of variational problem and whose a local
representative is $\phi : \mathrm{M} \to G/H$, each point $\phi(\mathrm{p})$ can be represented by
${\color{red}\phi_{\rm B}}(\mathrm p)$ and ${\color{blue}\phi_{\rm A}}(\mathrm p)$ which are completely determined by $\phi(\mathrm{p})$ and the structure of associated
bundle $\bigl( \mathrm{E}, \pi_{\mathrm{E}} , \mathrm{M} \bigr)$.
If we want to  treat type A and type B fields
as true dynamical degrees of freedom we need to express the effective Lagrangian
in terms of ${\color{red}\phi_{\rm B}}$ and ${\color{blue}\phi_{\rm A}}$, where 
${\color{red}\phi_{\rm B}}: \mathrm{M} \to G/\mathrm{Stab}(\bm \Sigma)$ and
${\color{blue}\phi_{\rm A}}: \mathrm{M} \to \mathrm{Stab}(\bm \Sigma)/H$ are
unknown functions, not specified by $\phi(\rm p)$. To do so, we first need
to reconstruct the section  $s: \mathrm{M} \to \mathrm{E}$.
Let   $({\color{red}z^{\alpha}}, {\color{blue}v^{\bar{\alpha}}})$, where
${\color{red}\alpha} = 1,2, \dots \mathrm{dim}\bigl(G/\mathrm{Stab}(\bm \Sigma)\bigr)$ and  ${\color{blue}\bar{\alpha}} = 1,2, \dots \mathrm{dim}\bigl(\mathrm{Stab}(\bm \Sigma)/H\bigr)$, be a local coordinate system on $G/H$ and let $(x^\mu, w^j)$, $j=1,2, \dots \mathrm{dim}\bigl(G\bigr)$
be a local coordinate system on $(\rm P, \pi_{\rm P}, \rm M)$. Then, the map $\phi: 
{\rm M} \to G/H$ can be expressed as
$\Bigl({\color{red}\phi_{\rm B}^\alpha},  {\color{blue}\phi_{\rm A}^{\bar \alpha}}    \Bigr)$,
where ${\color{red}\phi_{\rm B}^\alpha} = {\color{red}z^\alpha}\circ {\color{red}\phi_{\rm B}}$ and ${\color{blue}\phi_{\rm A}^{\bar \alpha}} = {\color{blue}v^{\bar \alpha}}\circ {\color{blue}\phi_{\rm A}}$.
Given $\phi(\rm p)$, we can obtain section $s$ using \eqref{SSectDef}. If
$\sigma^j = w^j\circ \sigma$, 
$\psi_\sigma:\pi_{\mathrm{P}}^{-1}(\mathrm{U}) \to \mathrm{U} \times G$ is a local trivialization
of $\pi_{\mathrm{P}}$ and $\psi_{\mathrm{E}}: f^{-1}({\color{red}\phi_{\mathrm{B}}}(\mathrm{U})) \to {\color{red}\phi_{\mathrm{B}}}(\mathrm U) \times 
\bigl(\mathrm{Stab}(\bm \Sigma)/H\bigr)$,
a local trivialization
on $G/H$, for each $\mathrm p \in \mathrm{U} \subset \mathrm{M}$ 
\begin{equation}
	s(\mathrm p) = q \Bigl( \psi_\sigma^{-1} \bigl( x^\mu(\mathrm p), \sigma^j(\rm p)\bigr), \psi_{\mathrm{E}}^{-1}\bigl({\color{red}\phi_{\rm B}^\alpha}(\rm p),  {\color{blue}\phi_{\rm A}^{\bar \alpha}}(\rm p) \bigr)  \Bigr)
\end{equation}
and effective Lagrangian can be obtained as discussed in Section \ref{NoetherJet}. Several
examples are given below.

\subsection{SO(3) ferromagnet and antiferromagnet} \label{HeisExample}

\noindent The construction from  previous subsections
enable us to compare two well known cases of spontaneous symmetry
breaking: SO(3) Heisenberg ferromagnet and antiferromagnet. Spontaneous symmetry breaking occurs along the scheme $\mathrm{SO}(3) \to \mathrm{SO}(2)$ in both of these
cases, yet  systems behave rather differently al low energies.
The mechanism which produces differences is well understood -- effective Lagrangians for both systems are derived in \cite{PRD}
and subsequent studies \cite{Hofmann1,Hofmann3,Hofmann8,PaperAnnPhys}
exploited them to reach at detailed description
of free field theories, possible magnon-magnon interactions, scattering processes and
low temperature thermodynamics in both cases \cite{Auerbach,NikolicPRB,PencoJHEP2024}. In short, the effective Lagrangian for an antiferromagnets takes a form of  Lorentz-invariant
nonlinear $\sigma$-model with the target space $S^2$ (type A NG fields) and non-interacting theory is of a Klein-Gordon
type. In contrast, the effective Lagrangian for a ferromagnet contains additional
term with single time derivative  of  magnon fields, frequently refereed to as a Wess-Zumino term, with fields also  taking values in $S^2$ (ferromagnetic magnons are classified as type B NG fields). The presence of this term, which makes the free magnon theory 
of a Schr\"{o}dinger type,  produces various
consequences in behavior of ferromagnets in comparison with antiferromagnets.

While magnon-magnon interactions are described in detail in both cases, as they follow from the effective Lagrangian, a minor
puzzle seems to remain to this day. Why do systems with the same SSB pattern
$\mathrm{SO}(3) \to \mathrm{SO}(2)$, and thus the same target space
$\mathrm{SO}(3) / \mathrm{SO}(2) = S^2$, behave differently?
According to analysis presented here, we see that the answer lies
in the existence of Lorentz-symmetry breaking order parameter -- spontaneous magnetization 
$\bm \Sigma \in \mathfrak{so}(3)^* \simeq \mathfrak {so}(3) \simeq \mathbb{R}^3$. The unbroken subgroup, $H = \mathrm{SO}(2)$, is in fact  the stabilizer $\mathrm{Stab}(\bm \Sigma) = \mathrm{SO}(2)$
and  $\mathrm{SO}(3)/ \mathrm{Stab}(\bm \Sigma) = S^2 $ is a two-sphere of radius $|\bm \Sigma|$ \cite{MarsdenSymmetry}. Thus, in the case of ferromagnets, the sphere $S^2$
is generated as a coadjoint orbit which comes equipped with canonical symplectic form, so the total number of physical degrees of freedom is one. Upon quantization, it becomes the ferromagnetic magnon which is classified as a type B Nambu-Goldstone boson \cite{JapanciPRL}.

\section{Coordinate expressions} \label{CoordExpres}

\noindent Coordinate-free approach presented in previous sections is useful for general discussions, but
explicit form of effective Lagrangian is needed for practical calculations. Since detailed accounts on coordinate expressions for various $G$-invariant term exist \cite{CCWZ1,CCWZ2,PRD,AnnPhys,PRX,JHEP,PRLJB}, we will focus here only on the one which is directly related to $\bm \Sigma$. Also, to compare 
results presented here with existing literature,
we shall assume $\mathrm{M} = \mathbb{R}^{m+1}$ so that $\pi$ is a trivial bundle and that $G$ describes an internal symmetry. Coordinate expressions also provide a direct illustration
of results obtained in previous section.%

\subsection{Coordinates on homogeneous space $G/H$}

\noindent
Suppose that the symmetry breaking occurs as described in
Section \ref{SecSymBr}. To discuss possible terms in effective
Lagrangian, we first need to specify the induced vector fields which
describe symmetry transformations on $G/H$ and corresponding jet bundle. If we choose 
$\{ u^a \}$ as standard coordinates on $G/H$, each point $\mathrm{a} \in G/H$
can be written as $\exp u \cdot \bm T$, where we use abbreviation
$u \cdot \bm T = u^a \bm T_a$ and $\{\bm T_a\}$ constitute a basis of $\mathfrak{m}$.
For given $\bm A \in \mathfrak{g}$, the induced field $X_{\bm A}$ at a
point $\mathrm{a} \in G/H$ is 
given by \cite{Helgason}
\begin{equation}
X_{\bm A}|_{\mathrm{a}} = \frac{\mathrm{d}}{\mathrm{d} \epsilon} \at[\bigg]{0} \Big{[}  \exp \epsilon \bm A 
\exp u \cdot \bm T \Big{]}_{\mathfrak{m}}  \equiv (X_{\bm A})^b \frac{\partial}{\partial u^b}\at[\bigg]{\mathrm{a}}
\end{equation}
where subscript $\mathfrak{m}$ denotes the restriction to $T_{\mathrm{a}}(G/H) \simeq \mathfrak{m}$ and we have used abbreviation $l(g,\mathrm{a}) = g\mathrm{a}$.
Now we specify a basis in $\mathfrak{g}$ so that
\begin{equation}
\Big{[} \bm T_i, \bm T_j \Big{]} = f_{ij}^{\;\;k} \bm T_k
\end{equation}
and use BCH formula to evaluate exponentials. As a result, we
obtain components of induced vector fields as a power series in coordinates on homogeneous space. The first few terms are
\begin{equation}
(X_{\bm A})^b = A^i \left[  \delta^b_i +  u^a f_{ia}^{\;\;k} 
\delta^b_k
+ \frac{1}{2} u^a u^c f_{ia}^{\;\;k} f_{ck}^{\;\; j} \delta^b_j + \dots \right]. \label{KillingVectSer}
\end{equation}
As efficient algorithms for evaluating BCS terms exist \cite{JMPBCH},
one can  calculate additional terms in a straightforward fashion.
The 1-jet prolongation of $X_{\bm A}$ is now given by \cite{Saunders}
\begin{equation}
X_{\bm A }^{(1)} = (X_{\bm A})^b\; \frac{\partial}{\partial u^b}
+ (X_{\bm A})^b_\mu \;\frac{\partial}{\partial u^b_\mu}
\end{equation}
where $u^b_\mu$ are derivative coordinates,
\begin{equation}
(X_{\bm A})^b_\mu  = \frac{\mathrm{d} (X_{\bm A})^b}{\mathrm{d} x^\mu}
\end{equation}
and 
\begin{equation}
\frac{\mathrm{d} }{\mathrm{d} x^\mu} = \frac{\partial}{\partial x^\mu}
+ u^b_\mu \frac{\partial}{\partial u^b}
\end{equation}
is the total derivative on $J^1 \pi$.

The Lorentz-symmetry breaking order parameter represents  the ground state expectation value of the Noether charge and thus it may be viewed as a 
$\mathfrak{g}^*$ valued timelike vector field $\bm \Sigma \otimes \partial_t$.  A term in effective Lagrangian, invariant up to total derivative [see \eqref{ActionInvDiv}] may be obtained by contracting
it with a pullback of  $\mathfrak{g}$ valued one-form on $G/H$. In 
terms of local coordinates on $G/H$, we have
\begin{eqnarray}
\left \langle \bm \Sigma \otimes \frac{\partial}{\partial t} , s^* c^i_a(u) \mathrm{d} u^a \otimes \bm T_i  \right \rangle
= \Sigma_i c_a^i(s) \frac{\partial \phi^a}{\partial t} 
= (j^1 s)^*c_a(u) \dot{u}^a \equiv (j^1s)^* L^{(0,1)} 
\end{eqnarray}
where $c_a(u) = \Sigma_i c_a^i(u)$, $\dot u^a = u^a_0$  and $\langle, \rangle$ denotes
natural paring between elements of $\mathfrak{g}$ and its dual, as well
as paring between tensor fields on $\mathrm{M}$.

According to \eqref{ActionInvDiv}, the invariance of classical field theory
corresponding to $L^{(0,1)}$ under the action of $G$ is expressed
as
\begin{equation}
X_{\bm A}^{(1)} \Big( L^{(0,1)}  \Big) \mathrm{Vol M} 
= \mathrm{d} \big( \Delta^\mu_{\bm A} \mathrm{Vol_\mu M} \big) \label{X1LieDer}
\end{equation}
Since
\begin{equation}
X_{\bm A}^{(1)} \Big( L^{(0,1)}  \Big) = \dot u^a X_{\bm A}^b \partial_b c_a + \dot{X}_{\bm A}^b c_b
= X_{\bm A}^b \big( \partial_b c_a - \partial_a c_b   \big) \dot{u}^a
+ \frac{\mathrm{d}}{\mathrm{d} t}\Big( X_{\bm A}^b c_b  \Big),
\end{equation}
by assuming only $\Delta^0_{\bm A} \neq 0$, and defining
$\Delta^0_{\bm A} = E_{\bm A} + X^a_{\bm A} c_a$, where $E_{\bm A}$
is a function on $\pi$ linear in $\bm A$, we obtain  equations which determine functions $c_a(u)$
\begin{equation}
X_{\bm A}^b \big( \partial_b c_a - \partial_a c_b   \big) = \partial_a E_{\bm A}. \label{CaEq}
\end{equation}
The components $X^b_{\bm A}$ are given in \eqref{KillingVectSer}, while
the meaning of functions $E_{\bm A}$ follows from \eqref{SigmaAverage}.
By letting $\bm A$ to be one of the generators of $\mathfrak{g}$, we see
that
\begin{equation}
\Sigma_i = \left \langle \frac{1}{V} \int_{\bar{\mathrm{U}}} s^* E_i  \; \mathrm{Vol_0M} \right \rangle_{\rm{g.s.}}
\end{equation}
and equation \eqref{CaEq} can be solved as described in \cite{PRD,PRX,JapanciPRL}.

Higher order terms in effective Lagrangian can be found in standard ways. For example, 
a term quadratic in derivatives of Nambu-Goldstone fields can be
constructed from unique $G$ invariant metric on $G/H$ built from the
$H$ invariant inner product on $\mathfrak{m}$ \cite{Grk}. By our assumption,
the Lorentz invariance is broken by $\bm \Sigma$, but if rotational symmetry remains,  the inverse Lorentz metric $\eta^{\mu \nu} \partial_\mu \otimes \partial_\nu$
splits into $\eta^{\mathrm{t}} \partial_t \otimes \partial_t$ and
$\eta^{\mathrm{s}} \delta^{rs} \partial_r \otimes \partial_s$ where
$\eta^{\mathrm{t}}$ and $\eta^{\mathrm{s}}$ are constants, $\eta^{\mathrm{s}}<0$ and
$r,s = 1,2, \dots m$. If
$g_{ab}\mathrm{d} u^a \otimes \mathrm{d} u^b$ is $G$ invariant metric on
$G/H$, two additional terms in effective Lagrangian are
\begin{eqnarray}
\left \langle \frac 12  \eta^{\mathrm{t}} \partial_t \otimes \partial_t, 
s^* g_{ab}\mathrm{d} u^a \otimes \mathrm{d} u^b  \right \rangle
&=&  \frac12 \eta^{\mathrm{t}}(j^1 s)^*g_{ab}(u) \dot{u}^a \dot{u}^b \equiv (j^1s)^* L^{(0,2)} ,  \nonumber \\
\left \langle \frac 12  \eta^{\mathrm{s}} \delta^{rs} \partial_r \otimes \partial_s, 
s^* g_{ab}\mathrm{d} u^a \otimes \mathrm{d} u^b  \right \rangle
&=&  \frac12 \eta^{\mathrm{s}}(j^1 s)^*g_{ab}(u) {u}^a_r {u}^b_r \equiv (j^1s)^* L^{(2,0)} .
\end{eqnarray}
The function $(j^1s)^*\left[  L^{(0,1)}+L^{(0,2)} +L^{(2,0)} \right]$
describes the dynamics of $\mathrm{dim}(G) - \mathrm{dim(H)}$ Nambu-Goldstone fields. To identify type B NG fields, one proceeds as described in
 \cite{JapanciPRL,PRX}. First, a quadratic approximation to Lagrangian is made after which one diagonalizes the term with a single time derivative
 which leads to the separation of type A and type B NG fields.

Closely related to $L^{(0,1)}$ is a wide class of Wess-Zumino terms -- contributions
to the effective action which are, under the action of $G$, invariant up to 
the surface terms. The one discussed here plays the dominant role  in
free field theory as it determines the dispersion relation of type B NG fields. 
WZ terms containing higher number of derivatives could be constructed in a similar
manner as $L^{(0.1)}$. This requires a careful treatment of differential forms and
de Rham cohomology of $G/H$, however \cite{BraunerWZW,DHoker}.

\subsection{Coordinates on bundle $G/H$} \label{CoordGmodH}

\subsubsection{Free field theory}

\noindent While, in general, local coordinates on $G/H$ mix two types of NG fields, a different view on space $G/H$ enables
us to separate type A and type B fields from the start. That is, with a suitable choice of coordinates on $G/H$, we can make distinction between
two types of NG fields in the effective Lagrangian more apparent.

As we noted in Section \ref{SecSymBr}, homogeneous space $G/H$ is in fact a bundle with base space $G/\mathrm{Stab}(\bm \Sigma)$ which is a symplectic manifold. The symplectic structure is provided by $G$  invariant Kirillov-Kostant (KK) symplectic form which is defined as follows. Let $\bm \Sigma \in \mathfrak{g}^*$ denote the Lorentz-symmetry breaking parameter, $\bm \Sigma' = \mathrm{Ad}^*(g)\bm \Sigma$ be another point on
$G/\mathrm{Stab}(\bm \Sigma)$ obtained from $\bm \Sigma$ by coadjoint action of $g \in G$ and let $X_{\bm A}$ and $X_{\bm B}$ be two induced vector fields on $G/\mathrm{Stab}(\bm \Sigma)$ corresponding to $\bm A, \bm B \in \mathfrak{g}$. The KK form
$\omega_{\bm \Sigma}$ at $\bm \Sigma'$ is defined as \cite{Kirillov}
\begin{equation}
\omega_{\bm \Sigma}(\bm \Sigma') \Big(X_{\bm A} \at[\big]{\bm \Sigma'}, X_{\bm B}\at[\big]{\bm \Sigma'}    \Big) := \bigl \langle \bm \Sigma',[\bm A, \bm B]   \bigr \rangle.
\end{equation}
We can now use  \eqref{Ndef} to decompose $\bm A$ and $\bm B$ as $\bm A  =  \bm A_{\mathfrak{stab}(\bm \Sigma)} + \bm A_{\mathfrak {n}}$
and $\bm B  =  \bm B_{\mathfrak{stab}(\bm \Sigma)} + \bm B_{\mathfrak {n}}$. Since $\bm \Sigma$ is $\mathfrak{stab}(\bm \Sigma)$
invariant, we have
\begin{equation}
\left \langle \bm \Sigma, [\bm A_{\mathfrak{stab}(\bm \Sigma)}, \bm B] \right \rangle
= \left \langle \bm \Sigma, \mathrm{ad}(\bm A_{\mathfrak{stab}(\bm \Sigma)}) \bm B
\right \rangle = - \left \langle
\mathrm{ad}^*(\bm A_{\mathfrak{stab}(\bm \Sigma)}) \bm \Sigma, \bm B
\right \rangle = 0,
\end{equation}
so that in the definition of $\omega_{\bm \Sigma}(\bm \Sigma)$ we may restrict
$\bm A$ and $\bm B$ to their components within $\mathfrak{n} = 
\mathfrak{g}/\mathfrak{stab}(\bm \Sigma)$.
If we pick a chart with coordinates $z^{\color{red}{\alpha}}$, where ${\color{red}{\alpha}}, {\color{red}{\beta}} = 1,2 \dots \mathrm{dim}(G/\mathrm{Stab}(\bm \Sigma))$, we can expand $\omega_{\bm \Sigma}$ as
\begin{equation}
\omega_{\bm \Sigma} = \frac12(\omega_{\bm \Sigma})_{{\color{red}{\alpha}} {\color{red}{\beta}}}\; \mathrm{d} z^{{\color{red}{\alpha}}} \wedge \mathrm{d} z^{{\color{red}{\beta}}}.
\end{equation}
and, to the lowest order, it is given by
\begin{equation}
\omega_{\bm \Sigma} =  \frac 12 \Sigma_i
f_{{\color{red}{\alpha \beta}}}^{\;\;\;\;i}\mathrm{d} z^{{\color{red}{\alpha}}} \wedge \mathrm{d} z^{{\color{red}{\beta}}}.
\end{equation}
Since $\mathrm{d} \omega_{\bm \Sigma} = 0$, we can write $\omega_{\bm \Sigma} = \mathrm{d} c_{\bm \Sigma}$, where
\begin{equation}
c_{\bm \Sigma}= \frac 12 \Sigma_i f_{{\color{red}{\alpha \beta}}}^{\;\;\;\;i} z^{{\color{red}{\alpha}}}\mathrm{d} z^{{\color{red}{\beta}}}
\end{equation}
 is an one-form on $G/\mathrm{Stab}(\bm \Sigma)$
which can be used to construct $L^{(0,1)}$
\begin{equation}
\left \langle  \frac{\partial}{\partial t} , s^* 
(c_{\bm \Sigma})_{{\color{red}{\beta}}}(z) \mathrm{d} z^{{\color{red}{\beta}}}  \right \rangle
= \frac 12 \Sigma_i f_{{\color{red}{\alpha}} {\color{red}{\beta}}}^{\;\;\;\;i} 
{\color{red}{\phi_{\mathrm{B}}^\alpha}} \partial_t
{\color{red}{\phi_{\mathrm{B}}^\beta}}
= L^{(0,1)}\bigl({{\color{red}{\phi_{\mathrm{B}}}}}, \partial_t {{\color{red}{\phi_{\mathrm{B}}}}}\bigr). \label{L01Bundle}
\end{equation}
It can be shown, 
 using arguments similar to those which lead to \eqref{CaEq}, that $L^{(0,1)}\bigl({{\color{red}{\phi_{\mathrm{B}}}}}, \partial_t {{\color{red}{\phi_{\mathrm{B}}}}}\bigr)$ changes by
a total derivative under the infinitesimal action of $G$.
We can now also choose a local chart on fibers over
$G/\mathrm{Stab}(\bm \Sigma)$. By denoting these coordinates  as
$v^{{\color{blue}{\bar{\alpha}}}}$, where ${{\color{blue}{\bar{\alpha}}}}, {{\color{blue}{\bar{\beta}}}} = 1,2 \dots \mathrm{dim}(\mathrm{Stab}(\bm \Sigma)/H)$,
we arrive at
the lowest order effective Lagrangian
\begin{equation}
L(\phi, \partial_t\phi, \partial_s\phi) = \frac 12 \Sigma_i f_{{\color{red}{\alpha}} {\color{red}{\beta}}}^{i} 
{\color{red}{\phi_{\mathrm{B}}^\alpha}} \partial_t
{\color{red}{\phi_{\mathrm{B}}^\beta}} - \frac12 (\eta^{\mathrm{s}}_{\mathrm{B}})_{{{\color{red}{\alpha}}} {{\color{red}{\beta}}}} \nabla  {\color{red}{\phi_{\mathrm{B}}^{{\color{red}{\alpha}}}}} \cdot 
\nabla  {\color{red}{\phi_{\mathrm{B}}^{{\color{red}{\beta}}}}} 
+ \frac 12 (\eta^{\mathrm{t}}_{\mathrm{A}})_{{{\color{blue}{\bar{\alpha}}}} {{\color{blue}{\bar{\beta}}}}}
\partial_t {\color{blue}{\phi_{\mathrm{A}}^{\bar{\alpha}}}}
\partial_t {\color{blue}{\phi_{\mathrm{A}}^{\bar{\beta}}}}
-\frac 12 (\eta^{\mathrm{s}}_{\mathrm{A}})_{{{\color{blue}{\bar{\alpha}}}} {{\color{blue}{\bar{\beta}}}}}
\nabla {\color{blue}{\phi_{\mathrm{A}}^{\bar{\alpha}}}} \cdot
\nabla {\color{blue}{\phi_{\mathrm{A}}^{\bar{\beta}}}}
+ O(\phi^4)
\end{equation}
where $(\eta^{\mathrm{s}}_{\mathrm{B}})_{{{\color{red}{\alpha}}}{{\color{red}{\beta}}}}$ are coefficients
defining a $\mathrm{Stab}(\bm \Sigma)$-invariant inner product on
$\mathfrak{g}/\mathfrak{stab}(\bm \Sigma)$, while 
$(\eta^{\mathrm{t}}_{\mathrm{A}})_{{{\color{blue}{\bar{\alpha}}}}{{\color{blue}{\bar{\beta}}}}}$ and
$(\eta^{\mathrm{s}}_{\mathrm{A}})_{{{\color{blue}{\bar{\alpha}}}}{{\color{blue}{\bar{\beta}}}}}$ are 
proportional to each other and define an $H$ invariant inner product on 
$\mathfrak{stab}(\bm \Sigma)/\mathfrak{h}$ \cite{FibrationEinstein,Besse}, and we have neglected
a term containing ${\partial_t\color{red}{\phi_{\mathrm{B}}^\alpha}} \partial_t
{\color{red}{\phi_{\mathrm{B}}^\beta}}$ \cite{PRX}.
This Lagrangian
describes the dynamics of noninteracting NG fields with type A and
type B fields  decoupled from each other. Further simplification can be achieved
with the use of Darboux coordinates around the origin in $G/\mathrm{Stab}(\bm \Sigma)$ in which $\omega_{\bm \Sigma}$ takes a diagonal form. This choice of coordinates may not be optimal for
other terms in the effective Lagrangian, however. Note also that when viewing $G/H$ as a bundle, $\bm \Sigma$ need not be explicitly included   in \eqref{L01Bundle} since the information
about the Lorentz-symmetry breaking order parameter is already 
encoded into space $G/\mathrm{Stab}(\bm \Sigma)$ via $\omega_{\bm \Sigma}$ and contraction of
$\phi^*c_{\bm \Sigma}$ with $\partial_t$ simply, from the point of view of effective Lagrangian, acknowledges that
Lorentz symmetry breaking did occur. Thus, even in the case when $G/H$ may be viewed
as an internal space (i.e. when the group $G$ describes an internal symmetry), the symplectic manifold 
$G/\mathrm{Stab}(\bm \Sigma)$ also reflects the properties of spacetime $\mathrm{M}$, namely the existence of a timelike vector field $\bm \Sigma \otimes \partial_t$ whose component is the vacuum expectation value of Noether charge.

\subsubsection{Nonlinear terms}

\noindent
We can go further with this construction and obtain nonlinear terms which,
in quantized theory, describe interactions among NG fields. In particular,
we shall identify the term responsible for leading order interaction between type A and type
B fields. The starting point is a well known result that $H$ invariant inner product
on $\mathfrak{m}$ induces a $G$ invariant metric on $G/H$ \cite{Besse,Grk}.

Suppose that SSB occurs in such a way that the order parameter $\bm \Sigma \in \mathfrak{g}^*$. According to  discussion from Section \ref{SecSymBr}, the Lie algebra
$\mathfrak{g}$ can be expressed as (See also \cite{Besse, FibrationEinstein})
\begin{equation}
	\mathfrak{g} = \mathfrak{h} \oplus \mathfrak{m}, \hspace{1cm} \mathfrak{m} = \mathfrak{n} \oplus \mathfrak{p}, \hspace{1.5cm}
\mathfrak{n} = 	\mathfrak{g}/\mathfrak{stab}({\bm \Sigma}), \;\;\;\;\; \mathfrak{p} = \mathfrak{stab}({\bm \Sigma})/\mathfrak{h},
\end{equation}	
and the inner product on $\mathfrak{m}$ decomposes as
\begin{equation}
	\langle\;\;, \;\; \rangle_{\mathfrak{m}} = \lambda_1 	\langle\;\;, \;\; 
	\rangle_{\mathfrak{g}/\mathfrak{stab}({\bm \Sigma})} \oplus 
	\lambda_2 	\langle\;\;, \;\; 
	\rangle_{\mathfrak{stab}({\bm \Sigma})/\mathfrak{h}}, \hspace{1cm} \lambda_1, \lambda_2 >0.
\end{equation}
This decomposition translates to the following form of metric tensor field on $G/H$
\begin{eqnarray}
	g_{G/H} & = &(g_{G/H})_{ab}(u) \mathrm{d} u^a \otimes \mathrm{d} u^b \nonumber \\
	& = & (g_{G/{\rm{Stab}}(\bm \Sigma)})_{{\color{red}\alpha} {\color{red}\beta}}(z) \mathrm{d} z^{\color{red}\alpha} \otimes \mathrm{d} z^{\color{red}\beta} \oplus 
	(g_{{\rm{Stab}}(\bm \Sigma)/H})_{{\color{blue}\bar{\alpha}} {\color{blue}\bar{\beta}}}(z,v) \mathrm{d} v^{{\color{blue}\bar{\alpha}}} \otimes \mathrm{d} v^{{\color{blue}\bar{\beta}}}
\end{eqnarray}
where $g_{G/{\rm{Stab}}(\bm \Sigma)}$ denotes the metric tensor field on symplectic
manifold $G/{\rm Stab}(\bm \Sigma)$ and $g_{{\rm{Stab}}(\bm \Sigma)/H}$ is the
corresponding field on the fiber ${\rm Stab}(\bm \Sigma)/H$. As before, 
$z^{\color{red}\alpha}$
denote local coordinates on $G/{\rm Stab}(\bm \Sigma)$ and $v^{{\color{blue}\bar{\alpha}}}$
are local coordinates on fibers. If $\{ \bm T_{{\color{red}\alpha}} \}$ denotes a basis of 
$\mathfrak{g}/\mathfrak{stab}({\bm \Sigma})$ and $\{ \bm T_{{\color{blue}\bar{\alpha}}} \}$ is
a basis of $\mathfrak{stab}({\bm \Sigma})/\mathfrak{h}$, each $\bm U = u^a \bm T_a \in \mathfrak{m}$ can be written as $\bm U = \bm Z + \bm V$, where $\bm Z = z^{\color{red}\alpha} \bm T_{\color{red}\alpha}$, $\bm V = v^{{\color{blue}\bar{\alpha}}}
\bm T_{{\color{blue}\bar{\alpha}}}$.
Matrix elements of $g_{G/H}$ can be calculated using inner products on $\mathfrak{g}/\mathfrak{stab}({\bm \Sigma})$ and $\mathfrak{stab}({\bm \Sigma})/\mathfrak{h}$ 
\cite{Muto}
\begin{equation}
(g_{G/{\rm{Stab}}(\bm \Sigma)})_{{\color{red}\alpha} {\color{red}\beta}}(z) = \left \langle 
\left[ \sum_{n=0}^\infty \frac{\left[ - {\rm ad} (\bm U)  \right]^n}{(n+1)!}
(\bm T_{{\color{red}\alpha}})  \right]_{\mathfrak{g}/\mathfrak{stab}({\bm \Sigma})}, 
\left[ \sum_{n=0}^\infty \frac{\left[ - {\rm ad} (\bm U)  \right]^n}{(n+1)!}
(\bm T_{{\color{red}\beta}})  \right]_{\mathfrak{g}/\mathfrak{stab}({\bm \Sigma})}\right \rangle,
\label{BaseMetric}
\end{equation}
\begin{equation}
(g_{{\rm{Stab}}(\bm \Sigma)}/H)_{{\color{blue}\bar{\alpha}} {\color{blue}\bar{\beta}}}(z,v) = \left \langle 
\left[ \sum_{n=0}^\infty \frac{\left[ - {\rm ad} (\bm Z + \bm V)  \right]^n}{(n+1)!}(\bm T_{{\color{blue}\bar{\alpha}}})  \right]_{\mathfrak{stab}({\bm \Sigma})/\mathfrak{h}}, 
\left[ \sum_{n=0}^\infty \frac{\left[ - {\rm ad} (\bm Z + \bm V)  \right]^n}{(n+1)!}
(\bm T_{{\color{blue}\bar{\beta}}})  \right]_{\mathfrak{stab}({\bm \Sigma})\mathfrak{h}}
\right \rangle \label{FiberMetric}
\end{equation}
By pulling $g_{G/{\rm{Stab}}(\bm \Sigma)}$ back
to spacetime, we see that $G/{\rm Stab}(\bm \Sigma)$ generates interactions
among type B NG fields. Additional nonlinear terms, which contain only ${{\color{red}{\phi_{\mathrm{B}}}}}$, arise from $s^* c_{\bm \Sigma}$. On the other hand, 
we see that the leading order interaction terms
which describe interaction among type A and type B NG fields are generated 
by the fiber metric $g_{{\rm{Stab}}(\bm \Sigma)/H}$ and this holds in general. Components
of higher order tensors, which play a role in perturbative calculations of quantized theory,  may be obtained in a similar fashion \cite{Grk}, but we will not analyze general
case here. Instead,  we shall focus on a specific example 
in Section \ref{SpinorBoseCOnd}.

\section{Explicit symmetry breaking} \label{ExplicitSSB}


\noindent
To study the effects of explicit symmetry breaking on the spectrum
of Nambu-Goldstone fields, imagine that we describe a   system  with
a $G$-invariant Hamiltonian $\mathsf{H}$. If the symmetry of the ground
state is $H \subset G$ invariant, and if it develops a $\mathrm{Stab}(\bm \Sigma)$-invariant Lorentz-symmetry breaking order
parameter $\bm \Sigma \in \mathfrak{g}^*$ as a vacuum expectation value of Noether charge,
the spectrum of Nambu-Goldstone fields contains both type A and type B
excitations [see(\ref{NPhiADef}) and (\ref{NPhiBDef})]. Suppose  that the 
symmetry of the model is explicitly broken by  finite charge density in 
the ground state. 
The system is now described by a modified Hamiltonian
$ \widetilde {\mathsf{H}} = \mathsf{H} - \mu^i Q_i$, where $\mu^i$ is the chemical
potential corresponding to the Noether charge $Q_i$ \cite{PRLMassive,SSPJHEP,PencoJHEP,HidakaPRD,GappedNREFT}. As we saw in Section \ref{NoetherJet},
the Noether charge is  an element of $\mathfrak{g}^*$, $\bm Q = Q_i \bm T^{*i}$,
so that $\mathsf{H}$ and $\widetilde{\mathsf{H}}$ differ by
$\left\langle \bm Q, \bm \mu \right \rangle$
where $\bm \mu = \mu^i \bm T_i$ is a constant element of $\mathfrak{g}$ corresponding
to the chemical potential. This observation allows us to identify $\widetilde{G}$, the symmetry group of $\widetilde{\mathsf{H}}$, as the stabilizer of $\bm Q$: $\widetilde{G} = \mathrm{Stab}(\bm Q)$ which we assume to be connected, simply connected Lie group. To account for subsequent spontaneous symmetry breakdown, we must take $H \subset \widetilde{G}$. 
Also, we shall assume that 
explicit symmetry breaking does not change 
the symmetry group of the ground state and,
with respect to the 
behavior of stabilizer $\mathrm{Stab}(\bm \Sigma)$, we distinguish between two interesting cases below. As we illustrate in examples following general discussion, our arguments also hold for theories defined using Lagrangian formalism where chemical potential $\bm \mu$ appears as the temporal component of the gauge field.

\subsection{\texorpdfstring{$\widetilde{\mathrm{Stab}(\bm \Sigma)} = {\mathrm{Stab}(\bm \Sigma)}$}{TEXT}}

\noindent
Consider first the situation in which,  with respect to the action of both $G$ and $\widetilde{G}$, $\mathrm{Stab}(\bm \Sigma)$ remains the same.
Since the spontaneous symmetry breaking pattern is  $\widetilde{G} \to H$,
the number of physical degrees of freedom carried by type A and type B NG fields
is given by
\begin{equation}
\widetilde N_{\mathrm{A}} = \mathrm{dim}\Bigl(  \mathrm{Stab}(\bm \Sigma)/H  \Bigr),
\hspace{1cm} \widetilde N_{\mathrm B} = \frac 12 \mathrm{dim} \biggl( \widetilde G/\mathrm{Stab}(\bm \Sigma)  \biggr) .  \label{MassNGF1}
\end{equation}
We see that the reduction of symmetry $G \to \widetilde{G}$ does not change
the number type A NG fields. On the other hand, the number of type B NG 
fields is reduced and
\begin{equation}
	N_{\mathrm{mNGF}}  = N_{\mathrm{B}} - \widetilde{N}_{\mathrm{B}} = \frac 12 \mathrm{dim} \biggl(
 G/\mathrm{Stab}(\bm \Sigma) \biggr)  - \frac 12 \mathrm{dim} \biggl( \widetilde G/\mathrm{Stab}(\bm \Sigma)  \biggr)
 \label{MassiveNGB}
\end{equation}
is the number of physical degrees of freedom carried by "massive" NG fields.
One way to understand \eqref{MassiveNGB} is to recall that, for $\widetilde{G} \to H$
symmetry breaking pattern, NG fields may be 
viewed as excitations generated 
by space-time depended $\widetilde{G}$ transformation of the ground state configuration
\cite{WeinbergQTF2}. All classical field configurations generated by elements $g\in G$ not contained
in $\widetilde{G}$ do not satisfy assumptions of the Goldstone theorem and thus
need not be "massless". That this is indeed the case may be showed, for example, by using effective Lagrangian and substitution $\partial_0 \phi^a\to \partial_0\phi^a -  (X_i)^a\mu^i$, together with additional terms containing no derivatives of NG fields \cite{PRLMassive}. In
particular, $L^{(0,1)}$ receives a contribution 
$s^* \left \langle  \bm \Sigma(z), \bm \mu     \right \rangle$ \cite{PRLMassive}, where
$\bm \Sigma(z) = \mathrm{Ad}^*(\exp z\cdot \bm T) \bm \Sigma$, $\bm T \in \mathfrak{n} \simeq \mathfrak{g}/\mathfrak{stab} (\bm \Sigma)$ and $z^\alpha$
are local coordinates on $G/\widetilde G$. Since $\mathrm{Ad}(g) \bm \mu = \bm \mu$, for
$g \in \widetilde G$, 
this term is $\widetilde G$ invariant and 
vanishes as $\mu_i \to 0$. The series expansion for $\Sigma_i(z)$
\begin{equation}
	\Sigma_i(z) = \Sigma_i + z^\alpha f_{\alpha i}^{\;\;k} \Sigma_k + \frac{1}{2}z^\alpha z^\beta f_{\alpha i}^{\;\;l}f_{\beta l}^{\;\;k} \Sigma_k + \dots
	\label{SigmaCoadjointSeries}
\end{equation}
which is useful for explicit calculations, can be obtained using the definition of coadjoint action.

The associated bundle which includes effects of explicit symmetry breaking may 
be build starting from the diagram
\begin{equation}\label{GroupFibr2}
	\begin{tikzcd}[row sep=huge]
		&
		G  \arrow[dl,swap,"\widetilde{\pi}_\mathrm{P}"] \arrow[dr,"\pi_{\tilde{\mathrm{P}}}"] &
		\\
		G/H \arrow[rr,"{f}"] \arrow[dr,"f'"',swap] && G/\mathrm{Stab}(\bm \Sigma) 
		\arrow[dl,swap, "\widetilde f"]\\
		 & G/ \widetilde{G}  &
	\end{tikzcd} 
\end{equation}
since the fiber of $\bigl(G/\mathrm{Stab}(\bm \Sigma) , \widetilde f,   G/\widetilde{G} \bigr)$
is $\widetilde{G}/\mathrm{Stab}(\bm \Sigma)$ and this is the target space 
for type B  excitations generated by SSB pattern $\widetilde{G} \to H$, while the 
total space $G/\mathrm{Stab}(\bm \Sigma)$ hosts  both massless and massive type
B NG fields. In other words, massive NG fields may be viewed as excitations
which take values on $G/\widetilde{G}$ which, since $G/\widetilde G$ is the
base space of bundle $G/\mathrm{Stab}(\bm \Sigma)$,
is a symplectic manifold carrying the KK form induced by $\bm \Sigma$. 
The number of physical degrees of
freedom  carried by massive NG fields is thus $(1/2)\mathrm{dim} \bigl( G/\widetilde{G} \bigr)$, in agreement with \eqref{MassiveNGB}. 
Also, the fiber of $\bigl(G/H, f',   G/\widetilde{G} \bigr)$ is
$\widetilde{G}/H$, which is the target space for all massless excitations in this case.
Further construction of associated
bundle  may be conducted  as outlined in Section \ref{NGFieldsAsocB}.

\subsection{\texorpdfstring{$\widetilde{\mathrm{Stab}(\bm \Sigma)} \subset {\mathrm{Stab}(\bm \Sigma)}$}{TEXT}}

\noindent
Suppose that, under the action of $\widetilde G$, the order parameter stabilizer
is reduced to  $\widetilde{\mathrm{Stab}(\bm \Sigma)} \subset \mathrm{Stab}(\bm \Sigma)$.
The numbers of physical
degrees of freedom corresponding to type A and type B NG fields are now
\begin{equation}
\widetilde N_{\mathrm{A}} = \mathrm{dim}\Bigl(  \widetilde{\mathrm{Stab}(\bm \Sigma)}/H  \Bigr),
\hspace{1cm} \widetilde N_{\mathrm B} = \frac 12 \mathrm{dim} \biggl( \widetilde G/\widetilde{\mathrm{Stab}(\bm \Sigma)}  \biggr) .  \label{MassNGF2}
\end{equation}
While the number of type A NG fields reduces, we see that none of  $N_{\mathrm{A}} =\mathrm{dim}\bigl(\mathrm{Stab}(\bm \Sigma)/H  \bigr)$ of  them
becomes massive. Thus, the number of 
physical degrees of freedom carried by
massive NG comes from the difference in type B NG  fields and  is given by
\begin{equation}
	N_{\mathrm{mNGF}}  = N_{\mathrm{B}} - \widetilde{N}_{\mathrm{B}} = \frac 12 \mathrm{dim} \biggl(
	G/\mathrm{Stab}(\bm \Sigma) \biggr)  - \frac 12 \mathrm{dim} \biggl( \widetilde G/\widetilde{\mathrm{Stab}(\bm \Sigma)}  \biggr)
	\label{MassiveNGB2}
\end{equation}
which obviously reduces to \eqref{MassiveNGB} if $\widetilde{\mathrm{Stab}(\bm \Sigma)} = \mathrm{Stab}(\bm \Sigma)$ and agrees with \cite{PRLMassive}.

To gain some additional insight into the formula \eqref{MassiveNGB2}, consider the following diagram
\begin{equation}
	\begin{tikzcd}[row sep=huge]
		&&
		G  \arrow[dll,swap,"\widetilde{\pi}_\mathrm{P}"] \arrow[drr,"\pi_{\tilde{\mathrm{P}}}"] \arrow[d] &&
		\\
		G/H \arrow[rr,"{f_1}"] \arrow[drr,"f'",swap]&& 
		G/\widetilde{\mathrm{Stab}(\bm \Sigma)}\arrow[rr,  "f_2"]
		\arrow[d,"f_3"] && G/\mathrm{Stab}(\bm \Sigma) \arrow[dll,"\widetilde f"]
		\\
		& &G/ \widetilde{G}  &
	\end{tikzcd}\label{GroupFibr3}.
\end{equation}
In comparison with \eqref{GroupFibr2}, we have three additional bundles: $f_1, f_2$
and $f_3$. The fiber of $f_1$ is $\widetilde{\mathrm{Stab}(\bm \Sigma)}/H$, which
is the target space for type A NG fields corresponding to $\widetilde{G} \to H$
spontaneous symmetry breakdown, and, as we have seen, they do not affect the
number of massive NG fields. On the other hand,  $G/\widetilde{\mathrm{Stab}(\bm \Sigma)}$ is the target space for type B NG fields
of $\widetilde{G} \to H$ as well as for massive NG fields.  
It may be viewed as the total space of both $f_2$ and $f_3$. Since respective fibers of 
these bundles are  $\mathrm{Stab}(\bm \Sigma)/\widetilde{\mathrm{Stab}(\bm \Sigma)}$ and $\widetilde{G}/\widetilde{\mathrm{Stab}(\bm \Sigma)}$, both of which are symplectic manifolds and represent two subspaces of $G/\widetilde{\mathrm{Stab}(\bm \Sigma)}$
which support massless type B NG fields, the number of physical degrees of freedom  carried by massive fields is
\begin{equation}
	N_{\mathrm{mNGF}} = \frac 12 \mathrm{dim}\biggl(
	G/\widetilde{\mathrm{Stab}(\bm \Sigma)}\biggr)   - \frac 12 \mathrm{dim} \biggl( \widetilde G/\widetilde{\mathrm{Stab}(\bm \Sigma)}\biggr) - \frac 12 \mathrm{dim} \biggl( \mathrm{Stab}(\bm \Sigma)/\widetilde{\mathrm{Stab}(\bm \Sigma)} \biggr)
	\label{MassiveNGB3}
\end{equation}
which is the same as \eqref{MassiveNGB2}. We can also add that, according to
\begin{equation}
	\begin{tikzcd}[row sep=huge]
&\widetilde{G}/H \arrow[r]& \widetilde G/\widetilde{\mathrm{Stab}(\bm \Sigma)}
\arrow[r]&
 \widetilde G/\mathrm{Stab}(\bm \Sigma)
\end{tikzcd}
\end{equation}
the type B NG fields separate into two additional categories: the ones which
take values on $\widetilde{G}/\mathrm{Stab}(\bm \Sigma)$ and the others
with target space $\mathrm{Stab}(\bm \Sigma)/\widetilde{\mathrm{Stab}(\bm \Sigma)}$.
We also note that
\begin{equation}
\widetilde{N}_{\rm A} + 2 \widetilde{N}_{\rm B} + 2 \widetilde{N}_{\rm{mNGF}}
= {\rm{dim}}\bigl(  G/H \bigr) - {\rm dim}\biggl( \mathrm{Stab}(\bm \Sigma)/\widetilde{\mathrm{Stab}(\bm \Sigma)} \biggr) = 
{\rm{dim}}\bigl(  G/H \bigr)-
\bigl(N_{{\rm A}} - \widetilde{N}_{{\rm A}}\bigr) \label{TotalCountGeneral}
\end{equation}
is the difference between  total number of fields in $G \to H$ and
$\widetilde G \to H$ SSB patterns when $\widetilde{\mathrm{Stab}(\bm \Sigma)} \subset {\mathrm{Stab}(\bm \Sigma)}$, since those NG fields of type A, which would be generated
by elements of $G$ in  $\mathrm{Stab}(\bm \Sigma)$, but not in $\widetilde{\mathrm{Stab}(\bm \Sigma)}$, disappear from the spectrum. Thus,
\begin{equation}
	N_{\rm{mNGF}} = \frac12 \biggl( G/\tilde{G} \biggr) - \frac 12 \biggl( \mathrm{Stab}(\bm \Sigma)/\widetilde{\mathrm{Stab}(\bm \Sigma)} \biggr)
\end{equation}
and we see that the target for space massive NG fields  is again a
symplectic manifold, $G/\widetilde{G}$, but their number is reduced when compared
to the case $\mathrm{Stab}(\bm \Sigma) = \widetilde{\mathrm{Stab}(\bm \Sigma)}$. 
Additional constraints can further decrease the number of massive NG fields (See Example \ref{SO2N}). The equation \eqref{TotalCountGeneral} is a general version
of $\widetilde{N}_{\rm A} + 2 \widetilde{N}_{\rm B} + 2 \widetilde{N}_{\rm{mNGF}}
= {\rm{dim}}\bigl(  G/H \bigr)$, which was obtained in \cite{Kapustin} by analyzing effective action. These relations are a direct consequence of fibrations given in \eqref{GroupFibr3}.
	



Finally, it may happen that the explicit symmetry breaking term changes
not only $\mathrm{Stab}(\bm \Sigma)$, but $H$ also. 
Depending on the relationship
between $G, \widetilde{G}, \mathrm{Stab}(\bm \Sigma), \widetilde{\mathrm{Stab}(\bm \Sigma)}, H$
and $\widetilde H$, where $\widetilde{H}$ denotes the symmetry group of the new ground state, 	one can identify massless and massive NG fields in each particular case using already established methods.

\section{Examples} \label{ExamplesSect}

\noindent
Here we illustrate general concepts with few examples. The aim of this section
is not to provide a detailed description of a particular physical system. Rather,
we want to demonstrate that coordinate-free method is not only suitable for general
analysis concerning numbers and of various NG fields, but that it can also be used
as a  basis for constructing specific effective Lagrangians.

\subsection{Heisenberg antiferromagnet}

\begin{figure}
	\centering
\includegraphics[width=\textwidth]{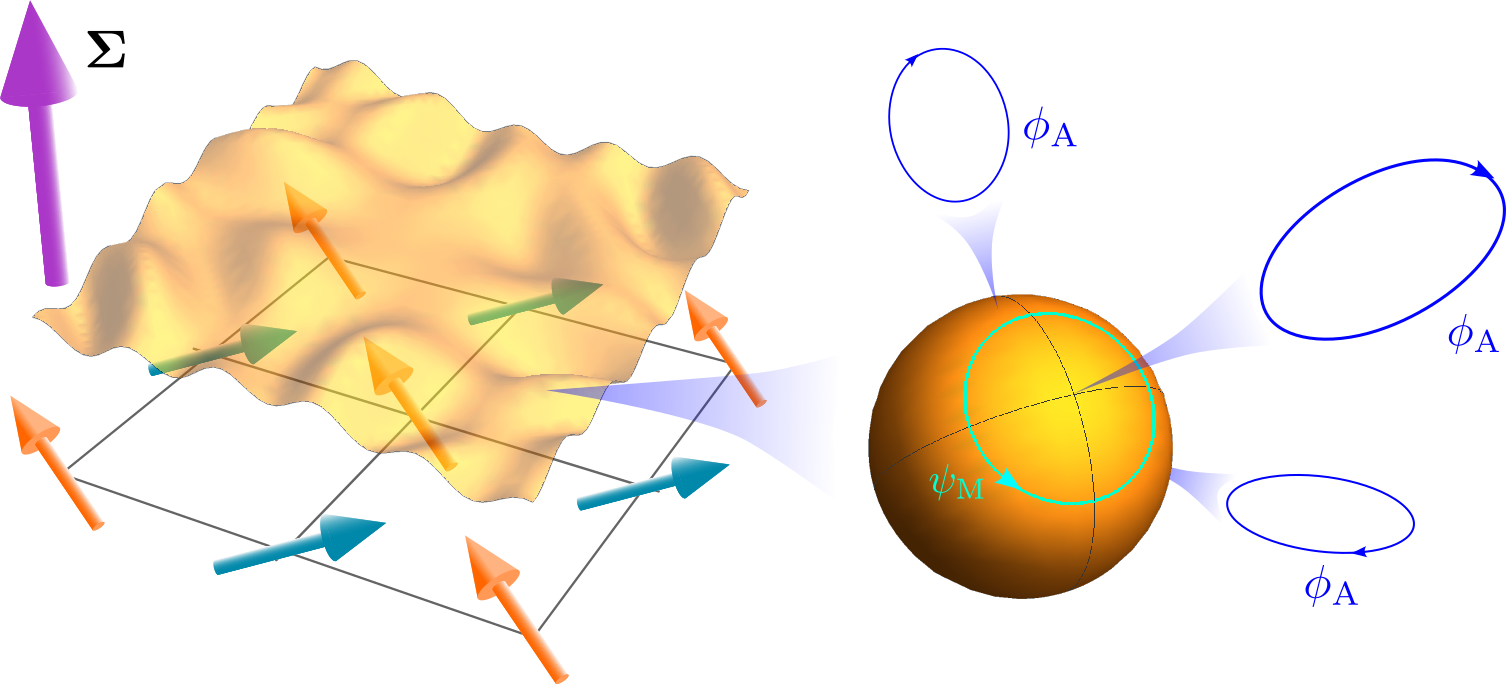} 
\caption{Canted spin configuration on a square lattice -- the symmetry of ground
state configuration is trivial, while emergent magnetization $\bm \Sigma$ is invariant under
the action of $\mathrm{SO}(2)$. Small deviations from the ground state are described by NG fields: Each point in the field space is a bundle so that massive NG fields  ${\color{Mgreen}\psi_{\mathrm M}}$ take values on sphere $S^2$, while type A NG field ${\color{blue}\phi_{\mathrm A}}$ lives on
$S^1$.}\label{HAFMfig}
\end{figure}

\noindent Heisenberg antiferromagnet is a famous example of $\mathrm{SO}(3) \to \mathrm{SO}(2)$ SSB. Due to the nature of exchange interaction, none of the generators
(components of total spin operator) develop vacuum expectation values and the spectrum
contains two NG fields of type A. The situation changes if the antiferromagnet
is put into  magnetic field $\bm B$. Suppose the field is oriented along the easy axis of 
antiferromagnetic system. This field explicitly breaks original $G = \mathrm{SO}(3)$
of Heisenberg Hamiltonian by reducing it to $\widetilde G = \mathrm{SO}(2)$. If the field is strong enough, the spins will rearrange themselves
in such a way that, at first, easy axis is perpendicular to magnetic field and, eventually, the canted phase will emerge \cite{AkhiezerAFM}. The canted
phase is characterized by complete breakdown of internal $\mathrm{SO}(3)$ symmetry and a non-zero value of net magnetization (See Figure \ref{HAFMfig}). 
Thus, $\mathrm{Stab}(\bm \Sigma) = \widetilde{\mathrm{Stab}(\bm \Sigma)} = \mathrm{SO}(2)$ and $H = \{ e\}$. Still, after SSB, there is
only one type A NG field and this is related to the fact that $\widetilde{G}= \mathrm{SO}(2)$ is Abelian. The spectrum also contains massive NG fields whose
number is determined by \eqref{MassiveNGB}.  This  gives $(3-1)/2 =1$ degree of freedom carried by massive NG field, which is a well known result
\cite{AkhiezerAFM,Callen,RomanSotoPRB,HofmannBrauner,PRLMassive}.

The lowest-order Lagrangian which describes the spin waves in this case may
be constructed as follows. Let ${\color{blue} v}$ denote the coordinate on $S^1$ so that
${\color{blue}\phi_{\mathrm A}} \equiv {\color{blue} v} \circ {\color{blue}\phi_{\mathrm A}}$ be a local representative of type A NG field, and let us choose coordinates on $S^2$ in such a way that  two massive NG
fields are collected into a complex field ${\color{Mgreen} \psi_{\rm M}} = 
\sqrt{|\bm \Sigma|/2}({\color{Mgreen} \phi_{\rm M}^1} + \mathrm{i} {\color{Mgreen} \phi_{\rm M}^2})$ \cite{AnnPhys2015}.
Then, to the order   $\bm p^2$, we have the following quadratic Lagrangian
\begin{equation}
	L = \frac{c_1}{2} \partial_t {\color{blue}\phi_{\mathrm A}} \partial_t {\color{blue}\phi_{\mathrm A}}
	-\frac{c_2}{2} \nabla  {\color{blue}\phi_{\mathrm A}} \cdot \nabla {\color{blue}\phi_{\mathrm A}} + \mathrm{i}  {\color{Mgreen} \psi_{\rm M}^*}
	\partial_t {\color{Mgreen} \psi_{\rm M}} - c_3 \nabla {\color{Mgreen} \psi_{\rm M}^*} \cdot \nabla {\color{Mgreen} \psi_{\rm M}} - B c_4 {\color{Mgreen} \psi_{\rm M}^*} {\color{Mgreen} \psi_{\rm M}}  \label{LagrHAFM}
\end{equation}
where $B$ denotes intensity of  external magnetic field $\bm B = B \bm T_3$
which plays the role of chemical potential. 
The therm proportional to magnetic field is $({\color{Mgreen} \psi_{\rm M}})^* \langle \bm \Sigma(z), \bm B\rangle$, with $\Sigma_i(z)$ expanded up to quadratic terms using \eqref{SigmaCoadjointSeries}.
Also, $c_1, c_2, c_3$ and $c_4$
are unspecified constants whose values may be determined by matching
with linear spin-wave theory or numerical simulations.

\subsection{Relativistic Bose-Einstein condensation}

\noindent
As a second example, consider a $G = \mathrm{O}(4) \simeq \mathrm{SU}(2) \times \mathrm{SU}(2)$ invariant theory of complex scalar doublet $\psi  =[\psi^1 \;\; \psi^2]^\mathrm{T}$ with standard Lagrangian
\cite{PRLMassive,Jakobsen}
\begin{equation}
	L = \partial_\alpha \psi^\dagger \partial^\alpha \psi -m^2 \psi^\dagger \psi- \lambda \bigl( \psi^\dagger \psi   \bigr)^2.
\end{equation}
Explicit symmetry breakdown in Lagrangian formalism is achieved by substitution
$\partial_0 \to D_0 =  \partial_0 -\mathrm{i} \mu$, where $\mu$ is the chemical potential, and
this reduces the symmetry of Lagrangian to  $\widetilde{G} = \mathrm{SU}(2) \times \mathrm{U}(1)$. If $\mu > m$, this model exhibits relativistic Bose-Einstein condensation and is used to study kaons in dense quark matter. We shall take commutation relations between elements of Lie algebra in this subsection to be $[\bm T_i, \bm T_j] = \mathrm{i} f_{ij}^{\;\;k} \bm T_k$.

The classical ground state of this model is $\psi_0 = [0 \;\;v]^{\mathrm{T}}$. To find its symmetry group, we can choose the basis for $\tilde{\mathfrak{g}} = \mathfrak{su}(2) \oplus \mathfrak{u}(1)$ as
$\{\bm T_i = \bm \sigma_i/2, \bm T_4 = I_{2 \times 2}/2 \}$, where $\bm \sigma_i$ are the Pauli matrices, and examine the action of an arbitrary element  of $\mathfrak{g}$
\begin{equation}
\bm A =\sum_{i = 1}^4 A^i \bm T_i
\end{equation}
on $\psi_0$.
The invariance condition $\bm A \psi_0 = 0$ yields $A^1=A^2 = 0$ and $A^3 = A^4$, leading to $\mathfrak{h} = \mathfrak{u}(1)$ and consequently $H = \mathrm{U}(1)$. Having determined $H$, we focus on the order parameter. Giving the fact that the Lagrangian contains the term with
single time derivative $\mathrm{i} \mu \left(\psi^\dagger \dot \psi - \dot \psi^\dagger \psi\right)$, and that under the action of $\bm A$ the field transforms ae $\delta \psi^B = \mathrm{i} A^k [\bm T_k]_C^{\;\; B}\psi^C \equiv \mathrm{i} A^k \mathcal{F}^B_k$, where $C,B = 1,2$
and $[\bm T_k]_C^{\;\;B}$ denoting matrix elements of generator $\bm T_k$, we find the  time component of conserved currents
\begin{equation}
J^0_k(x) = -\mathrm{i} \frac{\partial \mathcal{L}}{\partial \dot \psi^A(x)}\mathcal{F}^A_k(x)	= \mu
\psi^*_A(x)\bigl[  \bm T_k  \bigr]_B^{\;\;A} \psi^B(x) + \dots
\end{equation}
which, for the ground state configuration $\psi_0$, gives $J^0_1 = J^0_2 = 0$, 
$J^0_3 = - J^0_4 = - \mu v^2$. Therefore, the order parameter is given by
\begin{equation}
\bm \Sigma = \mu v^2 \biggl( - \bm T^{*3} + \bm T^{*4}  \biggr)
\end{equation}
To identify $\widetilde{\mathfrak{stab}(\bm \Sigma)}$, we start from the definition \eqref{StabSigmaAlg} and find all $\bm B \in \mathfrak{su}(2) \oplus \mathfrak{u}(1)$
such that, for arbitrary $\bm A \in \mathfrak{su}(2) \oplus \mathfrak{u}(1)$
the following holds
\begin{equation}
0 = 	\langle \mathrm{ad}^*(\bm B) \bm \Sigma, \bm A    \rangle
= \left\langle \bm \Sigma, -\left[\bm B, \bm A\right] \right\rangle = 
- \mathrm{i} f_{ij}^{\;\;k}B^i A^j \left \langle  \bm \Sigma, \bm T^k \right \rangle.
\end{equation}
which leads to $B^1 = B^2 =0$ and we get $\widetilde{\mathrm{Stab}(\bm \Sigma)} = \mathrm{U}(1) \times \mathrm{U}(1)$. By using similar procedure we find $\mathrm{Stab}(\bm \Sigma) = \widetilde{\mathrm{Stab}(\bm \Sigma)}$. Therefore, spectrum of this model
contains one massless NG field of type A, two massless NG fields of type B
(carrying one physical degree of freedom), as well as two massive NG fields (carrying one physical degree of freedom as well). This corresponds to one massless NG boson of type A, one of type B and one massive NG boson in quantized theory. Indeed, this is 
found by explicit calculations \cite{Jakobsen,PRLMassive,PRDHolographic}.

Effective Lagrangian for this model, which captures dynamics of non-interacting NG fields up to $\bm p^2$, may be constructed in a similar manner
as \eqref{LagrHAFM}. It is given by
\begin{equation}
	L = \frac{c_1}{2} \partial_t {\color{blue}\phi_{\mathrm A}} \partial_t {\color{blue}\phi_{\mathrm A}}
	-\frac{c_2}{2} \nabla  {\color{blue}\phi_{\mathrm A}} \cdot \nabla {\color{blue}\phi_{\mathrm A}} + \mathrm{i}  {\color{Mgreen} \psi_{\rm M}^*}
	\partial_t {\color{Mgreen} \psi_{\rm M}} - c_3 \nabla {\color{Mgreen} \psi_{\rm M}^*} \cdot \nabla {\color{Mgreen} \psi_{\rm M}} -  c_4 {\color{Mgreen} \psi_{\rm M}^*} {\color{Mgreen} \psi_{\rm M}}
	+\mathrm{i} c_5 {\color{red}\psi_{\rm B}^*}\partial_t{\color{red}\psi_{\rm B}}
	- c_6 \nabla {\color{red}\psi_{\rm B}^*} \cdot \nabla {\color{red}\psi_{\rm B}} \label{LagrSU2SU2}
\end{equation}
where $c_1 c_2, \dots c_6$ are arbitrary constants with $c_5 \propto \mu$ (As in previous example, we have absorbed $|\bm \Sigma| \propto \mu$ into the definition
of complex fields ${\color{red}\psi_{\mathrm{B}}}$ and ${\color{Mgreen} \psi_{\rm M}}$). This term also stems from $\langle \bm \Sigma(z), \bm \mu \rangle$ and 
can be obtained using \eqref{SigmaCoadjointSeries} and structure constants of 
$\mathfrak{su}(2) \oplus \mathfrak{u}(1)$, which vanish whenever $i,j$ or $k=4$.

\subsection{The $\mathrm{SO}(2N)$ vector model at fixed charge} \label{SO2N}

\noindent 
For our third example, consider a Lorentz-invariant model with additional $G = \mathrm{SO}(2 N)$ internal symmetry. In what follows, we shall use the fact that coadjoint orbits for  compact groups are equivalent with adjoint ones, where the identification between elements of $\mathfrak{g}$ and $\mathfrak{g}^*$ can be
done using an Ad-invariant  inner product on $\mathfrak{g}$ \cite{Kirillov}.
Suppose that the Lorentz symmetry, together with $\mathrm{SO}(2N)$ internal
symmetry, is explicitly broken by adding $-\mu \left( Q_1 +Q_2 + \dots + Q_k \right)$,
$k\leq N$
to Hamiltonian, with $Q_i$ may now be viewed as basis elements of $\mathfrak{so}(2N)$.
Since the same chemical potential is coupled to all $k$ generators, we find
\cite{Besse}
\begin{equation}
\widetilde G = \mathrm{Stab} \biggl( Q_1 + Q_2 + \dots + Q_k  \biggr)
= \mathrm{SO}(2N - 2k) \times\mathrm{U}(k)	.
\end{equation}
The $k$ fields transform under $U(k)$ and their configurations take part in subsequent spontaneous symmetry  breakdown while the remaining $2N-2k$ real field  are spectators.
Following \cite{AlvarezJHEP} we take the symmetry of ground state configuration to be 
$H = \mathrm{SO}(2N-2k) \times \mathrm{U}(k-1)$.
With Lorentz symmetry
being lost, one of the $k$ generators may develop a vacuum expectation value
$\Sigma \propto \langle Q_1 \rangle_{\mathrm{g.s.}} \neq 0$ and $\bm \Sigma = \langle Q_1 \rangle_{\mathrm{g.s.}} \bm T^{*1}$. 
Since   $\mathfrak{g}^*$ and $\mathfrak{g}$ have been identified, we have \cite{Besse}
\begin{equation}
	\mathrm{Stab}(\bm \Sigma) = \mathrm{Stab} \bigl( Q_1   \bigr) = \mathrm{SO}(2N - 2) \times\mathrm{U}(1)
\end{equation}
under the action of $G$. Likewise, the order parameter stabilizer for the action of $\mathrm{U}(k)$
is  $\mathrm{U}(k-1) \times\mathrm{U}(1)$ \cite{Grk}, thus
\begin{equation}
	\widetilde{\mathrm{Stab}(\bm \Sigma)} = \mathrm{SO}(2N - 2k) \times\mathrm{U}(k-1)
	\times\mathrm{U}(1).
\end{equation}
Having specified all relevant symmetry groups, we can analyze the spectrum of NG fields in this model for explicit $G \to \widetilde{G}$ and subsequent spontaneous symmetry breakdown $\widetilde{G} \to H$.

First, there is $\mathrm{dim}\left( \widetilde{\mathrm{Stab}(\bm \Sigma)}/ H  \right) = 1$ NG field of type A. Second, the number of physical degrees of freedom corresponding
to type B NG fields is given by \eqref{MassNGF2} and this yields $\widetilde{N}_{\mathrm B}=k-1$. Finally, the number of degrees of freedom carried by massive fields
can be read  off from \eqref{MassiveNGB2}, which gives $2 N -k-1$. However, 
 $2 N - 2 k$ of these massive modes are, by construction, spectator fields and thus  "frozen" under the action of full group $G$. Therefore, the number
 of massive NG fields is 
\begin{equation}
	N_{\mathrm{mNGF}} = k-1
\end{equation}
In other words, the NG sector of corresponding quantum theory contains one NG boson of type A, $k-1$ NG bosons of type B, $k-1$ massive NG bosons and the full spectrum also contains $2N - 2k$ massive particles
corresponding to
spectator fields.
The results presented here are in complete agreement with detailed calculations performed in \cite{AlvarezJHEP} (See also \cite{AlvarezPR,PRPLargeCharge}). As in the previous example, a direct
calculation reveals the presence of a one more massive degree of freedom corresponding
to the Higgs mode which is inaccessible to this construction.
The effective Lagrangian which describes dynamics of non-interacting NG fields to $\bm p^2$ in this case is
obtained as a direct generalization of \eqref{LagrSU2SU2}
\begin{eqnarray}
L & = & \frac{c_1}{2} \partial_\mu {\color{blue}\phi_{\mathrm A}} \partial^\mu {\color{blue}\phi_{\mathrm A}} + \frac{1}{2} \Sigma_1 f_{{\color{red}{\alpha}} {\color{red}{\beta}}}^{1} 
{\color{red}{\phi_{\mathrm{B}}^\alpha}} \partial_t
{\color{red}{\phi_{\mathrm{B}}^\beta}} - \frac{c_2}{2} \nabla {\color{red}\phi_{\rm B}^\alpha} \cdot \nabla {\color{red}\phi_{\rm B}^\alpha} \nonumber \\
& + & \frac{1}{2} \Sigma_1 f_{\color{Mgreen}\alpha \beta}^{1} 
{\color{Mgreen}{\phi_{\mathrm{M}}^\alpha}} \partial_t
{\color{Mgreen}{\phi_{\mathrm{M}}^\beta}} - \frac{c_3}{2}
\nabla {\color{Mgreen}\phi_{\rm M}^\alpha} \cdot \nabla {\color{Mgreen}\phi_{\rm M}^\alpha} + c_4 \frac{\mu^l}{2} \Sigma_1 {\color{Mgreen}{\phi_{\mathrm{M}}^\alpha}}
{\color{Mgreen}{\phi_{\mathrm{M}}^\beta}} f_{{\color{Mgreen}\alpha}l}^{\;\;j}
f_{{\color{Mgreen}\beta}j}^{\;\;1}
\end{eqnarray}
where $\alpha, \beta = 1,2, \dots k-1$ for all fields and $\mu_1 = \mu_2 \dots = \mu_k\equiv \mu$. 
Note that $G/\widetilde G$ is a subspace of $G/\widetilde{\mathrm{Stab}(\bm \Sigma)}$ and 
thus  inherits the symplectic structure defined by $\bm \Sigma$.
Had not we chosen $2N - 2k$
spectator fields, the spectrum would contain one NG boson of type A, $k-1$ NG bosons
of type B, $2N-k-1$ massive NG bosons and one Higgs boson. The total number of particles
is the same in both cases and equals $2N$.

\subsection{Spinor Bose-Einstein condensate} \label{SpinorBoseCOnd}

\noindent
As our final example, let us examine spinor Bose-Einstein condensate in the 
case of hyperfine spin $F=1$ \cite{SpinorBEC,JapanciPRL,TopologicalBEC}. 
The system is described by a three-component Schr\" odinger field $\bm \psi \in \mathbb{C}^3$ and a standard Lagrangian which is invariant under global $G = {\rm SO}(3) \times {\rm U(1)}$ transformations. Here, ${\rm SO}(3)$ acts as rotation on three-component field,
while ${\rm U(1)}$ transformation  corresponds to phase rotation of a complex field $\bm \psi$. 

Depending on the symmetry of the ground state, we distinguish between polar and ferromagnetic phases. Both of them share the same symmetry group of the ground state, $H= {\rm U}(1) \cong {\rm SO}(2)$, which is the diagonal subgroup of ${\rm SO}(3) \supset {\rm SO}(2) \times {\rm U}(1)$, and the ferromagnetic phase is characterized by the order parameter $\bm \Sigma \in \mathfrak{g}^* = \mathfrak{so}(3) \oplus \mathfrak{u}(1)$.
Thus,
\begin{equation}
	G/H = G = {\rm SO}(3) \times {\rm U(1)}/{\rm U}(1) \cong {\rm SO}(3)
\end{equation}
and there are three type A NG fields in the polar phase. The leading-order effective Lagrangian in this case is simply
\begin{equation}
	L = \frac{c_1}{2} (g_{{\rm SO}(3)})_{ab}({\color{blue} \phi_{\rm A}})\partial_t {\color{blue}\phi_{\mathrm A}^a} \partial_t {\color{blue}\phi_{\mathrm A}^b} - 
	\frac{c_2}{2} (g_{{\rm SO}(3)})_{ab}({\color{blue} \phi_{\rm A}}) \nabla {\color{blue}\phi_{\mathrm A}^a} \cdot \nabla {\color{blue}\phi_{\mathrm A}^b}
\end{equation}
where ${\color{blue} a} = 1,2,3$ and $g_{{\rm SO}(3)}$ is a $G$ invariant metric 
on the coset space ${\rm SO}(3)$.

Let us now move on more interesting ferromagnetic phase. If we choose a basis in $\mathfrak{g}^*$ so that $\bm \Sigma = \Sigma_3 \bm T^{*3} \equiv \Sigma_3 \bm T_3$, we have ${\rm Stab}(\bm \Sigma) = {\rm O}(2) \times {\rm U}(1)$. Thus, the symplectic manifold, which is the target space for NG fields of
type B is 
\begin{equation}
G/{\rm Stab}(\bm \Sigma) =  {\rm SO}(3) \times {\rm U(1)}/{\rm O}(2) \times {\rm U}(1) \cong S^2,
\end{equation}
while type A NG fields occupy fibers 
\begin{equation}
{\rm F} = {\rm Stab}(\bm \Sigma)/H =  {\rm O}(2) \times {\rm U}(1)/ {\rm U}(1) \cong 
{\rm O}(2) \cong {\rm U}(1).
\end{equation}
Therefore, the coset space $G/H $ is the bundle
\begin{equation}
  G/H = \Big{(} {\rm SO}(3), f, S^2, {\rm F} = {\rm U}(1)  \Big{)}
\end{equation}
where $f$ denotes the projection $f: {\rm SO}(3) \to S^2$, and the ferromagnetic phase is
characterized by single type B NG field and one type A NG field.

As we saw in section 
\ref{CoordGmodH}, the structure of $G/H$ not only determines the number of type A and
type B NG fields, but also provides us withe a direct instruction to construct effective Lagrangian which 
describes free fields, as well as  their interactions. To find lowest order nonlinear 
terms, we first decompose the Lie algebra $\mathfrak{g}$ as
\begin{equation}
\mathfrak{g}=	\mathfrak{so}(3) \oplus \mathfrak{u}(1) = \mathfrak{h} \oplus \mathfrak{n}
	\oplus \mathfrak{p} \equiv \mathfrak{h} + \mathfrak{m},
\end{equation}
where
\begin{equation}
  \mathfrak{u} = \mathfrak{u}(1), \hspace{1cm} \mathfrak{m} = \mathfrak{so}(3), 
	\hspace{1cm} \mathfrak{n} = \mathfrak{so}(3) \oplus \mathfrak{u}(1)/\mathfrak{u}(1)\oplus \mathfrak{o}(2)
	\cong \mathfrak{so}(3)/\mathfrak{o}(2), \hspace{1cm} \mathfrak{p} = \mathfrak{u}(1)\oplus \mathfrak{o}(2)/\mathfrak{u}(1) \cong \mathfrak{u}(1).
\end{equation}
Each $ \bm U = u^a \bm T_a \in \mathfrak{m}, \; a = 1,2,3,$ can be written as a sum $\bm U = \bm Z + \bm V$, where $\bm Z = z^{\color{red}\alpha} \bm T_{\color{red}\alpha},\; {\color{red}\alpha} = 1,2$ and $\bm V = v \bm T_{\color{blue}3}$. Now we can use
commutation relations $[T_{\color{red}\alpha}, T_{\color{red}\beta}] = \epsilon_{{\color{red}\alpha} {\color{red}\beta} {\color{red}\gamma}}T_{\color{red}\gamma}$
and \eqref{BaseMetric} to calculate matrix elements of the metric tensor on $g_{G/{\rm Stab}(\bm \Sigma)}$ as
\begin{equation}
	(g_{G/{\rm Stab}(\bm \Sigma)})_{{\color{red}\alpha} {\color{red}\beta}}(z) = 
	\delta_{{\color{red}\alpha} {\color{red}\beta}}
	 + \frac13 z^{\color{red}\mu} z^{\color{red}\nu} \epsilon_{{\color{red}\mu} 
	{\color{red}\alpha}{\color{red}\gamma}} \epsilon_{{\color{red}\beta} {\color{red}\nu}
	{\color{red}\gamma}} +
	 \mathcal{O}(z^3) = \delta_{{\color{red}\alpha} {\color{red}\beta}} + \frac13 \Big{[} z_{\color{red}\alpha} z_{\color{red}\beta} 
	 - \delta_{{\color{red}\alpha} {\color{red}\beta}}\big{(} (z_{\color{red}1})^2 +
	(z_{\color{red}2})^2 \big{)} \Big{]} +  \mathcal{O}(z^3).
\end{equation}
In a similar manner, we find
\begin{equation}
	(g_{{\rm Stab}(\bm \Sigma)/H})_{{\color{blue}33}}(z,v) = 1- \frac 13 \big{(}
	(z_{\color{red}1})^2 + (z_{\color{red}2})^2 \big{)}.
\end{equation}
Giving that the KK form does not generate lowest-order nonlinear terms in this case,
we find the effective Lagrangian which describes the ferromagnetic phase of  
$F=1$ spinor Bose-Einstein condensate
\begin{eqnarray}
	L({\color{blue} \phi_{\rm A}}, {\color{red} \phi_{\rm B}}) &=& \frac 12
	\Sigma_3 \Big{(} {\color{red} \phi_{\rm B}^1} \partial_t {\color{red} \phi_{\rm B}^2} - \partial_t {\color{red} \phi_{\rm B}^1} {\color{red} \phi_{\rm B}^2} \Big{)}
	-\frac 12 \eta_{{\color{red}\rm B}}^{\rm S} \nabla {\color{red} \phi_{\rm B}^\alpha}
	\cdot \nabla {\color{red} \phi_{\rm B}^\alpha} - \frac 16 
	\eta_{{\color{red}\rm B}}^{\rm S} 
	\Big{[} 
	{\color{red} \phi_{\rm B\alpha}} 
	{\color{red} \phi_{\rm B\beta}}-\delta_{{\color{red}\alpha}{\color{red}\beta}}  
	\big{(} ({\color{red} \phi_{\rm B}^1})^2 + ({\color{red} \phi_{\rm B}^2})^2  \big{)}
	\Big{]} 
	\nabla {\color{red} \phi_{\rm B}^\alpha}
	\cdot \nabla {\color{red} \phi_{\rm B}^\beta} \nonumber \\
	&+& \frac 12 \eta_{{\color{blue}\rm A}}^{\rm T} \left[ 
	1-\frac 13 
	\big{(} ({\color{red} \phi_{\rm B}^1})^2 + ({\color{red} \phi_{\rm B}^2})^2  \big{)}
	\right] \partial_t {\color{blue} \phi_{\rm A}} \partial_t {\color{blue} \phi_{\rm A}}
	- \frac 12 \eta_{{\color{blue}\rm A}}^{\rm S} \left[ 
	1-\frac 13 
	\big{(} ({\color{red} \phi_{\rm B}^1})^2 + ({\color{red} \phi_{\rm B}^2})^2  \big{)}
	\right] \nabla {\color{blue} \phi_{\rm A}} \cdot \nabla {\color{blue} \phi_{\rm A}}
	\label{SpinorBECLagr}
\end{eqnarray}
where we have neglected  terms containing $ \partial_t{\color{red} \phi_{\rm B}^\alpha} \partial_t {\color{red} \phi_{\rm B}^\beta}$ and $\eta_{{\color{red}\rm B}}^{\rm S}, \eta_{{\color{blue}\rm A}}^{\rm T}, \eta_{{\color{blue}\rm A}}^{\rm S}$ are arbitrary positive constants. As outlined in Section \ref{CoordGmodH}, interactions between type
A and type B NG fields arise from the fiber metric $g_{{\rm Stab}(\bm \Sigma)/H}$. In this example, the lowest-order interactions appear in the second line of 
\eqref{SpinorBECLagr}, while the remaining nonlinear terms describe interactions among
type B fields. 

Upon quantization,  Lagrangian  \eqref{SpinorBECLagr} describes interacting theory of a single type A NG boson and a single type B NG boson. We note that,  contrary to what is common practice \cite{SpinorBEC,JapanciPRL}, Lagrangian \eqref{SpinorBECLagr}
is constructed without any reference to the complex field $\bm \psi$. Because of that, perturbative calculations can be organized in a systematic way. All approximations concern only  interactions among physical degrees of freedom and eventual uncontrolled errors, which may arise due to  Hilbert space structure or commutation relations between operators of the original theory \cite{AnnPhys2015}, play no role here.




\section{Summary}

\noindent Nambu-Goldstone fields, the low energy degrees of freedom, play
prominent role in physical systems exhibiting spontaneous symmetry breakdown.
Their dynamics follows from the effective action which is the most general
$G$ invariant functional defined on an associated bundle whose
typical fiber is $G/H$. While the form of effective Lagrangian,
which involves local coordinates on $G/H$, is needed to derive equations
of motion and detailed prediction of theory based on perturbative
calculations, some important characteristics of NG fields
follow directly from the symmetry breaking pattern and the symmetry
of the order parameter $\bm \Sigma \in \mathfrak{g}^*$. 
A novel feature of the present work is precisely the coordinate-free formulation
of the first-order effective field theory which allows for a direct derivation of the number
of physical degrees of freedom   assigned to type A and type B Nambu-Goldstone
fields, including so-called massive NG fields.

The construction of classical field theory for type A and B NG fields
is implemented in several steps. By assuming symmetry breaking pattern
$G \to H$, we have shown that classical field theory in which NG fields
are maps $\phi: \mathrm{M} \to G/H$ could be cast in the language of jets 
where NG fields are identified as sections on associated bundle 
$(\mathrm{E}, \pi_{\mathrm{E}}, \mathrm{M})$ whose fibers
are $G/H$. Next, with nonzero Lorentz-symmetry breaking order parameter $\bm \Sigma \in \mathfrak{g}^*$ and 
$H \subset \mathrm{Stab}(\bm \Sigma)$, the coset
space $G/H$  becomes a bundle, the base space of which is the symplectic manifold 
$G/\mathrm{Stab}(\bm \Sigma)$ and the fibers are $\mathrm{Stab}(\bm \Sigma)/H$.
By using the structure of associated bundle, we demonstrated that to each field
configuration $\phi(\mathrm p) \in G/H$ corresponds a pair of points: 
${\color{red}\phi_{\mathrm{B}}}(\mathrm{p}) \in G/\mathrm{Stab}(\bm \Sigma)$
and  ${\color{blue}\phi_{\mathrm{A}}}(\mathrm{p}) \in \mathrm{Stab}(\bm \Sigma)/H$.
The maps  ${\color{red}\phi_{\mathrm{B}}}: \mathrm{M} \to G/\mathrm{Stab}(\bm \Sigma)$
and ${\color{blue}\phi_{\mathrm{A}}} : \mathrm{M} \to \mathrm{Stab}(\bm \Sigma)/H$
are local representatives of classical type B and type A NG fields and the number of physical degrees
of freedom carried by both types of fields is determined by $G, H$ and $\mathrm{Stab}(\bm \Sigma)$. Besides directly yielding  correct number of physical degrees of freedom,
this approach also provides a  path to Lagrangian  formulated in
terms of ${\color{blue}\phi_{\mathrm{A}}}$ and ${\color{red}\phi_{\mathrm{B}}}$.
We also examine the case of explicit symmetry breaking by finite charge density
in the ground state which reduces the symmetry of model from $G$ to $\widetilde{G}$. By
adapting arguments from first four sections to this case, 
we demonstrated how effective field theory in geometric formulation
may be used to predict the number of degrees of freedom corresponding to
type A, type B and massive NG fields, latter of  which are defined to be field configurations obtained by the the action
of elements of $G$ not contained in $\widetilde G$.
As the target space for massive NG fields is $G/\widetilde{G}$, the
base space of bundle $G/\mathrm{Stab}(\bm \Sigma)$, massive
NG fields  also couple through (the pullback of) KK symplectic form generated by $\bm \Sigma$.
In this sense, they turn out to be quite similar to type B NG fields. 
It is also possible that, under the action of $\widetilde G$, the isotropy group $\mathrm{Stab}(\bm \Sigma)$
reduces to $\widetilde{\mathrm{Stab}(\bm \Sigma)}$. The effect of this change 
is the reduction in the number of massive NG fields
 by $\mathrm{Stab}(\bm \Sigma)/\widetilde{\mathrm{Stab}(\bm \Sigma)}$.
Finally, we illustrate
general ideas presented here  with several examples. We stress once more that the classification
of NG fields presented here is completely determined by $G$, $H$, ${\rm Stab}(\bm \Sigma)$, $\widetilde{G}$ and $\widetilde{{\rm Stab}(\bm \Sigma)}$ and that it does not
depend on Lagrangian function. In particular, none of the  
procedures which aim at obtaining an approximate expression for the
dispersion relation of NG fields, with the help of which the classification is performed,  are used here.
This makes the method presented in the paper far more robust than the standard ones.

The 
existence of Lorentz-symmetry breaking order parameter $\bm \Sigma \in \mathfrak{g}^*$, which induces 
symplectic structure on $G/\mathrm{Stab}(\bm \Sigma)$, is a question which cannot be completely resolved within classical field theory discussed here and needs to be addressed
in corresponding quantum or statistical theory. The transition to quantized theory would allow for 
detailed perturbative studies and
geometric methods presented here could also be of use when analyzing the effects
of interactions between NG bosons.
All results presented in this paper are unrelated to a choice of local coordinates on $G/H$, and, by extension, the Lagrangian function, as well as to a choice of basis in $\mathfrak{g}$, which makes this formulation general and applicable to various physical
situations ranging from high energy particle physics all the way to condensed matter
and statistical systems.

\begin{acknowledgments}
	\noindent
	The author would like to thank professor Julio Guerrero for
	useful discussions.
	The author also acknowledges the generous hospitality of
	the Department of Mathematics at the University of Ja\'en 
	during the spring semester of 2024.
\end{acknowledgments}

\appendix

\section{Jet bundles and variation principle} \label{AppA}

\noindent
We collect here some basic facts about jet bundles. The exposition 
 supplements Section \ref{NoetherJet},  while the details can be
found in \cite{Saunders,KMS,GaugeJetBook,Momentum,SymmetriesLFT}.

A fibered manifold consists of two spaces $\rm E$ and  $\rm M$
and a surjective submersion $\pi_{\rm E}: \rm E \to \rm M$. In this context, the spaces $\rm E$ and $\rm M$ are known as the total space and the base space, respectively, while
the map $\pi_{\rm E}$ is known as the projection.  Also, the set $\pi_{\rm E}^{-1}(\rm p)$, where $\rm p \in \rm M$, is called the  fiber over $\rm p$ and is denoted
by $\rm E_{\rm p}$. Therefore, the triple $(\rm E, \pi_{\rm E}, \rm M)$ denotes a fibered manifold, although we sometimes refer to it simply as  $\pi_{\rm E}$. 
A local trivialization of $\pi_{\rm E}$ around $\rm p \in M$ is the triple $({\rm W}_{\rm p}, {\rm F}_{\rm p}, t_{\rm p})$, where ${\rm W}_{\rm p}$ is a neighborhood of $\rm p$, ${\rm F}_{\rm p}$
is a space and $t_{\rm p}$ is a diffeomorphism $t_{\rm p}: \pi_{\rm E}^{-1}({\rm W}_{\rm p}) \to {\rm W}_{\rm p} \times {\rm F}_{\rm p}$ for which ${\rm pr}_1 \circ \pi_{\rm E} = \pi_{\rm E} |_{\pi_{\rm E}^{-1}({\rm W}_{\rm p})}$ holds, with $\rm pr_1$ denoting the projection onto the first factor. A bundle is a fibered manifold which
is locally trivial, meaning that it has at least one local trivialization around
each of its points. It turns out that for bundles, all spaces $\rm F_{\rm p}$ are 
diffeomorphic to a typical fiber $\rm F$ \cite{Saunders}.
In the present paper, $\rm M$ denotes spacetime and we take its dimension
to be $m+1$, while we assume that the dimension of the typical fiber to be $n$. In the setting of effective field theory
we shall be dealing with typical fibers which are homogeneous spaces ${\rm F} = G/H$.
In this case, the bundle $(\rm E, \pi_{\rm E}, \rm M)$ can be constructed as an associated bundle \cite{KMS}.

Given a local trivialization, one can think of a bundle to be locally (with respect to the base space $\rm M$) given by product of $\rm W_{p}$ (a neighborhood of $\rm p \in M$)
and the typical fiber $\rm F$. We are thus lead to so called adapted coordinate systems.
The coordinates of each point $\rm a \in E$, with $\pi_{\rm E}(\rm a) = \rm p$, are given by $y(\rm a)$, where
$y = (x^\mu \circ {\rm pr_1} \circ t_{\rm p}, u^a \circ {\rm pr_2} \circ t_{\rm p})$
and $\mu = 1,2, \dots, m+1, a= 1,2, \dots, n$. It is common to write adapted coordinates simply as $(x^\mu, u^a)$.

A local section of the bundle $\pi_{\rm E}$ is a map, $s: \rm M \to E$, for which $\pi_{\rm E} \circ s = {\rm id}_{\rm M}$. Here ${\rm id}_{\rm M}$ denotes the identity 
function on $\rm M$. Since $s(\rm p) \in E$, we can use coordinates to represent a local section $s$ as $(x^\mu, s^a)$, with $s^a = u^a \circ s$. We denote by $\Gamma_{\rm p}(\pi_{E})$ the set of sections whose domains contains the point $\rm p \in M$.

Suppose we are given two bundles $(\rm E, \pi_{\rm E}, \rm M)$ and $(\rm H, \pi_{\rm H}, \rm N)$.
A function which maps the fibers of $\pi_{\rm E}$ into the fibers of $\pi_{\rm H}$ is a bundle
morphism. A bundle morphism is a pair of maps $(\eta, \bar \eta)$, where $\eta: \rm E \to \rm H$, 
$\bar \eta: \rm M \to \rm N$ and the condition $\pi_{\rm H} \circ \eta = \bar \eta \circ \pi_{\rm E}$.
Commutative diagrams such as
\begin{equation} 
	\begin{tikzcd}[row sep=1.8cm, column sep=2.3cm]
		& {\rm E} \arrow[r, "\eta"] \arrow[d, "\pi_{\rm E}"'] & {\rm H} \arrow[d, "\pi_{\rm H}"] \\
		 & \mathrm{M} 
		 \arrow[r, "\bar \eta"] 
		 & \mathrm{N} &
	\end{tikzcd}
\end{equation}
are often used to illustrate bundle morphisms. If $s \in \Gamma_{\rm W}(\pi_{\rm E})$,
with $\rm W \subset \rm M$,
and $\bar \eta$ is a diffeomorphism, we can define $s' \in \Gamma_{\bar \eta(\rm W)}(\pi_{\rm H})$
by $s':= \eta \circ s \circ \bar \eta^{-1}$. The relation between $s$ and $s'$ is
shown in the diagram below
\begin{equation} 
	\begin{tikzcd}[row sep=1.8cm, column sep=2.3cm]
		& {\rm E} \arrow[r, "\eta"] \arrow[d, "\pi_{\rm E}"] & {\rm H} \arrow[d, "\pi_{\rm H}"'] \\
		 & \mathrm{M} \arrow[u, "s", bend left]
		 \arrow[r, "\bar \eta"] 
		 & \mathrm{N}\arrow[u, "s'"', bend right] &
	\end{tikzcd}.
\end{equation}
Of particular importance is the case $\rm E = H, \rm M = N$ when the pair $(\eta, \bar \eta)$ describes automorphism of the bundle $\pi_{\rm E}$.

Jets of sections are defined as equivalence relations. We say that two sections $s$ and
$s'$ are 1-equivalent at $\rm p \in M$ if $s, s' \in \Gamma_{\rm p}(\pi_{E})$, $s({\rm p}) = s'(\rm p)$
and
\begin{equation}
	\frac{\partial s^a}{\partial x^\mu}\at[\bigg]{\rm p}  =  \frac{\partial s'^a}{\partial x^\mu}\at[\bigg]{\rm p}
\end{equation}
for all $a = 1,2, \dots, n$ and $\mu = 1,2,\dots, m+1$. This equivalence class is
called 1-jet of $s$ at $\rm p \in M$ and is denoted by $j^1_{\rm p}s$. It can be shown
that the definition of this equivalence relation does not depend on local
coordinates on $\rm E$. The set of all first order jets of local sections of $\pi_{\rm E}$ possesses
a structure of a differentiable manifold. The manifold is known as the
first jet manifold of $\pi_{\rm E}$ and
is defined as the set $J^1\pi_{\rm E}:= \{ j^1_{\rm p}s| \;{\rm p \in M}, s \in \Gamma_{\rm p}(\pi_{\rm E})\}$. Two surjective submersions can be defined on the manifold $J^1 \pi_{\rm E}$, and they are related to so-called induced coordinates. If $y= (x^\mu, u^a)$
represent adapted coordinates on (a portion of) $\rm E$, the induced coordinates 
on (a portion of) $J^1\pi_{\rm E}$ are $y^1 = (x^\mu, u^a, u^a_\mu)$ and are
defined as
\begin{equation}
	x^{\mu}(j^1_{\rm p}s) := x^{\mu}({\rm p}), \;\;\;\;\;\;
	u^a (j^1_{\rm p}s) := u^{a}\big{(}s({\rm p})\big{)}, \;\;\;\;\;\;
	u^a_\mu (j^1_{\rm p}s) := \frac{\partial s^a}{\partial x^\mu}\at[\bigg]{\rm p}.
\end{equation}
The first surjective submersion is $\pi_{1,0}: J^1 \pi_{\rm E} \to \rm E$, and is defined as $\pi_{1,0}(j^1_{\pi_{\rm E}(\rm a)}s) = \rm a = s\big{(}\pi_{\rm E}(\rm p)\big{)}$, or by $\pi_{1,0} \circ y^1 = y$. Therefore, the triple $(J^1 \pi_{\rm E}, \pi_{1,0}, \rm E)$ defines a bundle which
turns out to be an affine bundle. The second projection, $\pi_1 : J^1 \pi_{\rm E} \to \rm M$ is defined by $\pi_{1}(j^1_{\rm p}s) = \rm p$,  or by $\pi_1 := \pi_{\rm E} \circ \pi_{1,0}$, and the triple $(J^1\pi_{\rm E}, \pi_1, \rm M)$ also constitutes a bundle. Let $s \in \Gamma_{\rm p}(\pi_{\rm E})$. Then
$j^1s \in \Gamma_{\rm p}(\pi_1)$ is the
first prolongation of $s$, defined by $j^1s({\rm p}):= j^1_{\rm p}s$. 
The Lagrangian density is a function $L: J^1\pi_{\rm E} \to \mathbb{R}$
and the sections
of $\pi_1$ appear in variational problems of first order field theories. The action of 
a field theory is expressed as
\begin{equation}
	S[s]=\int_{\rm U}(j^1s)^* \mathcal L  \label{Sj1sDef}
\end{equation}
where $\mathrm{U} \subset \mathrm M$ is a compact submanifold of $\rm M$, $\mathcal L = L \rm{Vol} M$ is the Lagrangian form and $\rm Vol M$ is a volume form on $\rm M$.

A bundle morphism $(\eta, \bar \eta)$ can be prolonged to $j^1(\eta, \bar \eta)$
which maps the fibers of $J^1\pi_{\rm E}$ into the fibers of $J^1\pi_{\rm H}$. A special
case of bundle automorphism is discussed in Section \ref{NoetherJet}.
Let us parametrize the bundle automorphism $\eta$ by a real parameter $\epsilon \in \mathrm I$, where $\rm I$ is some interval containing $\epsilon = 0$ and $\eta_\epsilon = \eta$ for $\epsilon = 0$. The variation of a section $s$ is now defined in terms of $\epsilon$ as $s_\epsilon = \eta_\epsilon \circ s \circ (\bar \eta_\epsilon)^{-1}$, where
$\bar \eta_\epsilon$ is a diffeomorphism for all $\epsilon$. Consequently, [see \eqref{j1fDef}] the induced variation of $j^1s$ is given by $j^1s_\epsilon = j^1\eta_\epsilon \circ j^1s \circ (\bar \eta_\epsilon)^{-1}$ and we define 
\begin{equation}
	S_\epsilon[s]:= \int_{\bar \eta_\epsilon(\rm U)} (j^1s_\epsilon)^*\mathcal{L}
\end{equation}
to be the variation of the original action \eqref{Sj1sDef}. The action is characterized
as invariant under the automorphism $\eta$ if
\begin{equation}
	\frac{\mathrm{d}}{\mathrm{d} \epsilon} \at[\bigg]{0} \int_{\bar{\eta}_\epsilon
		(\mathrm U)} 
	(j^1s_\epsilon)^* \mathcal{L} = 0 \label{Invarj1SDef}.
\end{equation}
whether or not the equations of motion for $s$ hold. Note that, due to the fact that
$\bar \eta_{\epsilon}$ is nontrivial, the domain of integration in \eqref{Invarj1SDef}
is also affected by the bundle automorphism $(\eta_\epsilon, \bar \eta_\epsilon)$. 

\section{Projectable vector fields and invariance of the action} \label{AppB}

\noindent
Vector fields on ${\rm E}$ are sections of the tangent bundle $(T{\rm E}, \tau_{\rm E}, \rm E)$. The space of all vector fields on $\rm E$ is commonly denoted as
$\chi(\rm E)$. In adapted coordinates, an arbitrary vector field $X \in \chi(\rm E)$ takes
a form \cite{Saunders}
\begin{equation}
	X^\mu(x^\mu, u^a) \frac{\partial}{\partial x^\mu} + X^a(x^\mu, u^a) \frac{\partial}{\partial u^a}
\end{equation}
so that all the components are functions of the adapted coordinates. A vector field $X \in \chi(\rm E)$ may play a role of  a bundle morphism
from  $(\rm E, \pi_{E}, M)$ to $(T{\rm E}, \pi_{\rm E*}, T{\rm M})$, where
$\pi_{\rm E*}$ denotes the differential of the projection $\pi_{\rm E}$
\begin{equation} 
	\begin{tikzcd}[row sep=1.8cm, column sep=2.3cm]
		& {\rm E} \arrow[r, "X"] \arrow[d, "\pi_{\rm E}"'] & T{\rm E} \arrow[d, "\pi_{\rm E*}"]  \arrow[r, "\tau_{\rm E}"] & E \arrow[d,"\pi_{\rm E}"] &\\
		& \mathrm{M} 
		\arrow[r, "\bar X"] 
		& T\mathrm{M} \arrow[r, "\tau_{\rm M}"] & M &
	\end{tikzcd}
\end{equation}
This kind of a vector field is known as a projectable vector field. We see
from the diagram that $\tau_{\rm M}\circ \bar X \circ \pi_{\mathrm{E}} = \tau_{\rm M} \circ \pi_{\mathrm{E} *} \circ X = \pi_{\rm E} \circ \tau_{\rm E} \circ X = \pi_{\mathrm{E}}$, since $X$ is a section
of $\pi_{\rm E}$. Thus, $\tau_{\rm M}\circ \bar X = {\rm id}_{\rm M}$ and $\bar X$ is a section of $\tau_{\rm M}$ (that is, $\bar X$
must be a vector field on $\rm M$) and it is completely determined by $\bar X \circ \pi_{\mathrm{E}} = \pi_{\mathrm{E} *} \circ X$. This means that a projectable
vector field, in general, has a coordinate expression
\begin{equation}
		X^\mu(x^\mu) \frac{\partial}{\partial x^\mu} + X^a(x^\mu, u^a) \frac{\partial}{\partial u^a}.
\end{equation}
Flows of projectable vector fields are bundle morphisms. In other words, if $\eta_\epsilon$ denotes a bundle morphism parametrized by $\epsilon$, corresponding
projectable vector field is given by \eqref{ProjXDef} and similarly for $\bar X$.
Just as bundle morphisms, vector fields can be prolonged so that $X^1$ defined 
in \eqref{X1Jet} is the generator of $j^1\eta$ \cite{SCFT,Saunders,Momentum}. We can now express the invariance
condition in terms of generators (i.e. projectable vector field). Since
\begin{equation}
	\int_{\bar{\eta}_\epsilon (\mathrm U)} (j^1s_\epsilon)^* \mathcal{L} =
	 \int_{\mathrm U}(\bar{\eta}_\epsilon)^* (j^1s_\epsilon)^* \mathcal{L} = 
	 \int_{\mathrm U} (j^1s_\epsilon \circ \bar{\eta}_\epsilon)^* \mathcal{L},
\end{equation}
and $j^1s_\epsilon = j^1 \eta_\epsilon \circ j^1 s \circ (\bar \eta_{\epsilon})^{-1}$, we have
\begin{equation}
	0 = \frac{\mathrm{d}}{\mathrm{d} \epsilon} \at[\bigg]{0} \int_{\bar{\eta}_\epsilon
		(\mathrm U)} 
	(j^1s_\epsilon)^* \mathcal{L} = \frac{\mathrm{d}}{\mathrm{d} \epsilon} \at[\bigg]{0} \int_{\rm U}(j^1\eta_{\epsilon}\circ j^1 s)^* \mathcal{L}
	= \int_{\rm U} (j^1s)^* \frac{\mathrm{d}}{\mathrm{d} \epsilon} \at[\bigg]{0} 
	(j^1 \eta_\epsilon)^* \mathcal{L} = \int_{\rm U} (j^1s)^* {\rm d}_{X^1} \mathcal{L}
\end{equation}
where ${\rm d}_{X^1}$ denotes the Lie derivative with respect to the generator of $j^1 \eta$.
This results allows us to express the invariance condition directly in terms of the Lagrangian
form. Implications of this equation, in the case when typical fiber is a homogeneous space, are explored in detail in the main body of the paper.

\bibliography{Refs}

\end{document}